\documentclass[a4paper,11pt]{article}
\usepackage[utf8]{inputenc}
\usepackage[english]{babel}
\usepackage{amsmath,amsfonts,amssymb,amsthm,mathtools}
\usepackage{graphicx}
\usepackage{float}
\usepackage{cmap}
\usepackage{mathtext}
\usepackage{physics}
\usepackage{feynmp-auto}
\graphicspath{{pictures/}}
\DeclareGraphicsExtensions{.pdf,.png,.jpg,}
\usepackage[left=2cm,right=2cm,
    top=2cm,bottom=2cm,bindingoffset=0cm]{geometry}
\usepackage{subcaption}
\usepackage{caption}
\DeclareCaptionLabelFormat{cont}{#1~#2\alph{ContinuedFloat}}
\usepackage[nottoc]{tocbibind}

\usepackage{cite}
\usepackage{hyperref}

\title{Importance of unitarity in the search for heavy neutrinos in $e^+ e^- \to W^+ W^-$}
\author{G. A. Chachava, S. I. Godunov}
\date{\today}

\begin{document}

\maketitle

\begin{abstract}
   We study the process $e^+e^- \to W^+W^-$ with the aim of estimating the prospects for observing heavy neutrino contributions at future $e^+e^-$-colliders. In this work, we consider two implementations of heavy-light neutrino mixing: a linearized mixing approximation applied in popular models and an exact unitary mixing scheme. We conclude that the approximate realization leads to physically incorrect results for this process, while exact unitary mixing provides some signatures that can be experimentally checked.
\end{abstract}

\section{Introduction}

The process $e^+e^-\to W^+W^-$ is well known and included in many textbooks. For instance, the cancellation between $s$ and $t$ channels, needed for $S$-matrix unitarity~\cite{Vainshtein:1971ip,Weinberg:1971fb,Bogomolnyi:1973nj,Okun:ITEPWS1}, requires the existence of a Higgs boson with the electron coupling given by the electron mass~\cite{LQ}. In~\cite{Peskin:1995ev}, it is demonstrated how this cancellation follows from the Ward identities. There are various plans to build new lepton colliders in the future, and it is a good reason for us to revisit this beautiful physics.

In this study, we investigate the process $e^+e^- \to W^+W^-$, with particular emphasis on the possible contributions of hypothetical Heavy Neutral Leptons (HNL). Numerous experimental searches for HNLs have been conducted in the past, resulting in various constraints on their masses and mixing parameters. In the large HNL mass region, the strongest limits come from the LHC~\cite{CMS:2023jqi,CMS:2024hik,CMS:2024xdq,CMS:2024ake,ATLAS:2024fcs}. 

One of the most popular theoretical frameworks incorporating heavy neutrinos is the seesaw type\nobreakdash-I model \cite{Minkowski:1977sc,Gell-Mann:1979vob,Yanagida:1979as,Mohapatra:1979ia}. The main feature of this model is a natural explanation of the smallness of light neutrino masses. It is worth noting that it also allows for sizable active–sterile mixing through the introduction of additional heavy neutrino states \cite{Asaka:2015oia}. 

The influence of heavy neutral leptons on the differential cross section of the $e^+e^- \to W^+W^-$ process has been recently analyzed in \cite{Drutskoy:2024hqx} in the context of future $e^+e^-$-colliders. That paper was inspired by the seesaw type-I model, however, the general case was considered without specifying a theoretical model. In the present work, we perform a systematic study of this process within the seesaw type-I framework and identify the regions in the parameter space --- spanning heavy neutrino masses and mixing angles --- where their contribution becomes non-negligible or even dominant at certain center-of-mass energies. 

The key idea of our article is to compare two approaches to the implementation of heavy neutrinos in the Standard Model. The first one is to add them unitarily, which makes the Pontecorvo-Maki-Nakagawa-Sakata (PMNS)-mixing matrix nonunitary. The second one is to add them linearly while maintaining the  PMNS-matrix unitarity. The latter is implemented in the popular \texttt{HeavyN} model~\cite{Pascoli:2018heg}, which was utilized in~\cite{Drutskoy:2024hqx}. In this article, we compare these two approaches in the context of the process $e^+e^- \to W^+W^-$ and analyze the key features of each approach, emphasizing the implications for the HNL experimental search. For the exactly unitary mixing we provide the \texttt{FeynRules}~\cite{Christensen:2008py,Alloul:2013bka} model that can be used with packages for both analytical and numerical calculations. We use \texttt{MadGraph}~\cite{Alwall:2014hca} for numerical calculations in the SM seesaw type-I extension.

Though we always consider $e^+e^-$ collisions, muon colliders could be regarded in complete analogy with the only difference being the present bounds on heavy neutrino mixing parameters. The relevant projects are CEPC~\cite{CEPCStudyGroup:2023quu} ($HZ$), FCC-ee~\cite{FCC:2025lpp} ($t\bar{t}$),\footnote{The $HZ$ mode yields comparable constraints at both CEPC and FCC-ee. Therefore, we do not consider both colliders here, but analyze this mode for CEPC, where it benefits from a slightly higher integrated luminosity. For the $t\bar{t}$ mode, FCC-ee provides a higher integrated luminosity; accordingly, this mode is studied for FCC-ee.} ILC~\cite{ILCInternationalDevelopmentTeam:2022izu}, CLIC~\cite{CLICdp:2018cto}, and muon colliders~\cite{InternationalMuonCollider:2025sys}. The set of corresponding energies and luminosities is presented in Table~\ref{tab:colliders}. 
This defines the ranges of invariant mass $\sqrt{s}$ and heavy neutrino masses considered in the text.

\begin{table}[t]
\centering
\begin{tabular}{|l|cccccc|}
\hline
 Project & CEPC ($HZ$) & FCC-ee ($t\bar{t}$) & ILC & CLIC & BM1 & BM2 \\
\hline
$\sqrt{s}$ [GeV] & 240 & 365 & 1000 & 3000 & 3000 & 10000 \\
$\mathcal{L}$ [ab$^{-1}$] & $13.0$ & $2.7$ & $8.0$ & $5.0$ & $1.0$ & $10.0$ \\
\hline
\end{tabular}
\caption{Collider configurations used in the analysis}
\label{tab:colliders}
\end{table}

Section~\ref{sec:model} provides a concise overview of the theoretical foundations of the seesaw type-I mechanism and summarizes the key relations among its parameters. In Sec~\ref{Toy_Model_sec}, we analyze the simplified model of the $e^+e^- \to W^+W^-$ process with a single heavy neutrino and examine the behavior of the cross section for different parameters of the model. In Sec.~\ref{See_Saw_exten}, we compare our ``toy model'' and SM seesaw type-I extension and analyze the behavior of the cross section when three heavy neutrinos are introduced. In Sec.~\ref{bounds}, we perform the evaluation of possible constraints on the mixing parameters of heavy neutrinos that can be obtained at future lepton colliders. Section~\ref{concl} is devoted to summarizing all our results and formulating the main conclusions of the study.

\section{Seesaw Type-I}
\label{sec:model}

In this section, we follow the presentation given in~\cite{ParticleDataGroup:2024cfk}, Ch. 14. Throughout this work, all neutrino fields are assumed to be Majorana particles.

Within the framework of the seesaw type-I model, $\mathcal{N}$ heavy neutrino states are introduced as singlets under SM gauge symmetries. These heavy states mix with the massless neutrinos, so after mass matrix diagonalization, we obtain three light and $\mathcal{N}$ heavy neutrinos. The components of the weak doublets are linear combinations of mass eigenstates, e.g., the electron neutrino is given by:
\begin{equation}
    \nu_{Le} = P_L\qty(\sum_{i=1}^3 U_{ei}\nu_i + \sum_{I=1}^\mathcal{N} V_{eI} N_I),
\end{equation}
where $P_L=\frac{1+\gamma_5}{2}$ is the left\footnote{Let us note that $\gamma_5$ is defined according to~\cite{LL4, LQ}.} projector, $U_{ei}$ are the elements of the PMNS-mixing matrix, and $V_{eI}$ are heavy neutrinos' mixing parameters. 

Let us consider constraints on mixing parameters and masses:
\begin{enumerate}
   \item In this model, the masses and mixing parameters of both light and heavy neutrinos are constrained by the so-called seesaw relation:
   \begin{equation}\label{seesaw}
     m_{\text{eff}}^\nu + \sum_{i=1}^\mathcal{N} V_{eI}^2 M_I = 0,
   \end{equation}
   where $m_{\text{eff}}^\nu = \sum\limits_{i=1}^3 U_{ei}^2 m_i$. Here $m_i$ and $M_I$ are masses of light and heavy neutrinos, respectively. This relation imposes specific limits on the possible values of the mixing parameters, protecting the Lagrangian from mass terms that violate SM gauge symmetries, e.g., $m\bar{\nu}_{Le}\nu_{Le}^c$.

   \item Searches for neutrinoless double beta decay yield a lower bound on the half-life period that depends on the mixing of heavy neutrinos and the masses of neutrinos: 
   \begin{equation}\label{half}
     T_{1/2}^{-1} = \mathcal{A}\frac{m_p^2}{\langle p^2 \rangle}|m_{\text{eff}}|^2,
   \end{equation}
   where $m_p$ is the proton mass, $\mathcal{A}$ is the amplitude of a specific decay (see \cite{Faessler:2014kka} for details), and
   \begin{equation}
      \label{eff}
       m_\text{eff} = m_{\text{eff}}^\nu + \sum\limits_{I=1}^\mathcal{N} V_{eI}^2 M_If_I.
   \end{equation}
   Here, $f_I = \frac{\langle p^2 \rangle}{\langle p^2 \rangle+M_I^2}$; $\sqrt{\langle p^2 \rangle} \approx 200$ MeV \cite{PhysRevLett.125.252502}. The lower bound for the half-life period is an upper bound for the effective neutrino mass (\ref{eff}). Modern bounds on the effective neutrino mass are the following [see \cite{PhysRevLett.125.252502} and \cite{KamLAND-Zen:2024eml} respectively]:

\begin{align}
   \label{eq:UB_Gerda}
   |m_\text{eff}| & < m_\text{eff}^{\text{UB}}  = (0.079 - 0.180)\hspace{0.1cm} \text{eV}, \\
   \label{eq:UB_KZ}
   |m_\text{eff}| & < m_\text{eff}^{\text{UB}}  = (0.028 - 0.122)\hspace{0.1cm} \text{eV}.
\end{align}
At the same time, an upper bound on $|m_{\text{eff}}^\nu|$ can be taken as follows \cite{Esteban:2024eli}:
\begin{equation}
   \label{eq:nufit}
   |m_{\text{eff}}^\nu| \lesssim 0.1 \hspace{0.1cm} \text{eV}.
\end{equation}

Upper bounds~\eqref{eq:UB_Gerda},~\eqref{eq:UB_KZ}, and~\eqref{eq:nufit} in combination with the definition~\eqref{eff} constrain the quantities $V_{eI}^2$.

\item There are bounds arising from the electroweak precision tests (EW PO) \cite{Fernandez-Martinez:2016lgt}: 
\begin{equation}
    \sum\limits_{I=1}^{\mathcal{N}}|V_{eI}|^2 \equiv |V_{eN}|^2 < 1.25\cdot 10^{-3}.
   \label{eq:bounds_abs}
\end{equation}
However, since this bound is indirect and strongly depends on the impact of additional New Physics (of much heavier scale) contributions to loop corrections affecting particle masses and decay widths, and it can therefore be relaxed. In light of this, the strongest bound from EW~PO comes from the invisible $Z$-boson width and gives $\abs{V_{eN}}^2 \lesssim 2\cdot 10^{-2}$. The most stringent direct constraints have been obtained at the LHC~\cite{ATLAS:2024rzi,CMS:2024xdq}. In the following, \emph{solely for illustrative purposes} of strong mixing, we will use the mixing parameter $\abs{V_{eN}}^2 = 2.25\cdot 10^{-2}$ from~\cite{Drutskoy:2024hqx}. This will allow us to compare our results with the ones obtained in~\cite{Drutskoy:2024hqx}.
\end{enumerate}

In \cite{Asaka:2015oia}, it has been demonstrated that all these constraints, except \eqref{eq:bounds_abs}, can be alleviated if three heavy neutrinos are introduced.

Let us note that, in the general case, mixing involving three heavy neutrinos is described by a relatively large number of parameters, which introduces a certain degree of arbitrariness into the theory. However, in order for heavy neutrinos to be observable in collider experiments, they must exhibit sizable mixing parameters. At the same time, achieving large mixing typically requires a cancellation, which can occur only for specific choices of the model parameters. Such a mechanism was proposed in~\cite{Asaka:2015oia}, and we will adopt this framework in what follows. It is designed in a way that only contribution from one of neutrinos is relevant in the broad range of energies. Therefore, it should be equivalent to single HNL with large mixing. That is why we consider this scheme as a rather general case. 

Overall, we will take the following list of bounds on mixing parameters and masses of neutrinos that our theory should satisfy:
\begin{align}
    &\phantom{{} = {}|} m_{\text{eff}}^\nu + \sum_{i=1}^\mathcal{N} V_{eI}^2 M_I = 0,\\
    |m_\text{eff}| & = \abs{m_{\text{eff}}^\nu + \sum\limits_{I=1}^\mathcal{N} V_{eI}^2 M_If_I} < 0.180\hspace{0.1cm} \text{eV}, \\
    |m_{\text{eff}}^\nu| & = \abs{\sum\limits_{i=1}^3 U_{ei}^2 m_i} < 0.1 \hspace{0.1cm} \text{eV}, \\
    |V_{eN}|^2 & = \sum\limits_{I=1}^\mathcal{N}|V_{eI}|^2 < 1.25\cdot 10^{-3}.
\end{align}

However, it was shown in \cite{Asaka:2015oia} that, in the case of $\mathcal{N}=3$ and for a hierarchical mass spectrum $M_{N_1} \ll M_{N_2} \ll M_{N_3}$, the mixing parameters of $N_1$ and $N_3$ can be made negligibly small, leaving only one dominant contribution to $e^-e^- \to W^-W^-$ from a single heavy neutrino $N_2$. Such cancellation usually appears due to some approximate symmetry in the theory \cite{Kersten:2007vk}. However, since our study examines a different process, we should check if the resulting cross sections are dominated by a single heavy neutrino (see Sec.~\ref{See_Saw_exten}).

The incorporation of heavy neutrinos into the Standard Model can be carried out in two ways\nobreakdash --- exactly or within a linear approximation. In the first approach, heavy neutrinos are introduced in a fully unitary manner, satisfying the relation:
\begin{equation}
    \sum_{i=1}^3|U_{ei}|^2+\sum_{I=1}^{\mathcal{N}}|V_{eI}|^2 = 1.
\end{equation}
It should be noted that under this construction, the matrix $U$ ceases to be strictly unitary, which, in turn, modifies the coupling constants of the light neutrinos.

The second approach introduces heavy neutrinos with mixing parameters $V_{eI}$, while keeping the matrix $U$ unitary:
\begin{equation}
    \sum_{i=1}^3|U_{ei}|^2 + \sum_{I=1}^{\mathcal{N}}|V_{eI}|^2 = 
    1 + \sum_{I=1}^{\mathcal{N}}|V_{eI}|^2 = 1 + \abs{V_{eN}}^2.
\end{equation}
Though strictly speaking it is a dangerous thing to do, since the neutrino kinetic term would not be in the canonical form in either the flavor or mass basis, for many processes it is a reasonable approximation since the value of $\abs{V_{eN}}^2$ is small; see~\eqref{eq:bounds_abs}. This latter method is the one implemented in the \texttt{HeavyN} model~\cite{Pascoli:2018heg}, and it provides adequate results for a broad class of processes. However, it turns out that in certain cases such an approximate treatment leads to physically inconsistent results (i.e., violation of $S$-matrix unitarity), in which case the fully unitary approach must be employed. An important example of this is the process $e^+e^- \to W^+W^-$, which will be analyzed in what follows.

\section{Toy Model}
\label{Toy_Model_sec}

The analytical calculation of the process $e^+e^- \to W^+W^-$ in the Standard Model is rather bulky due to the large number of terms contributing to the amplitude (see Sec.~\ref{See_Saw_exten} for details). In this section, we consider a simplified model --- a gauge theory with only an $SU(2)$ symmetry group containing gauge bosons $W$ and $Z$ of equal mass $M_Z=M_W\equiv M$, and a massless electron. Due to the fact that heavy neutrinos contribute only to diagrams with left-handed leptons and right-handed antileptons, we also consider all electrons to be left-handed particles and positrons to be right-handed particles throughout the paper. Within this ``toy model'', an analytic result is much simpler since the process involves only two diagrams: the $s$-channel with a $Z$-boson exchange and the $t$-channel with an electron neutrino exchange (see Fig.~\ref{fig_toy}). Despite its simplicity, this model retains all the essential features present in the Standard Model. It could be understood in the following way: a diagram with a Higgs boson compensates all effects from electron mass \cite{LQ}, and $\gamma$ and $Z$ diagrams interfere negatively, neutralizing the effects of electroweak mixing \cite{Peskin:1995ev}.
    \begin{figure}[t]
    \centering
        \begin{fmffile}{cross21}
            \begin{fmfgraph*}(180,120)
                \fmfleft{i1,i2}
                \fmfright{o1,o2}
                \fmf{fermion}{v1,i2}
                \fmf{fermion}{i1,v1}
                \fmf{boson, label=$Z$}{v1,v2}
                \fmf{boson}{v2,o2}
                \fmf{boson}{v2,o1}
                \fmflabel{$e^+$}{i2}
                \fmflabel{$e^-$}{i1}
                \fmflabel{$W^-$}{o1}
                \fmflabel{$W^+$}{o2}
            \end{fmfgraph*}
        \end{fmffile}
        \hspace{0.7 cm}
        \begin{fmffile}{cross31}
            \begin{fmfgraph*}(180,120)  
                \fmfleft{i1,i2}
                \fmfright{o1,o2}
                \fmf{photon}{v2,o2}
                \fmf{photon}{o1,v1}
                \fmf{fermion, label=$\nu_e$}{v1,v2}
                \fmf{fermion}{i1,v1}
                \fmf{fermion}{v2,i2}
                \fmflabel{$e^+$}{i2}
                \fmflabel{$e^-$}{i1}
                \fmflabel{$W^-$}{o1}
                \fmflabel{$W^+$}{o2}
            \end{fmfgraph*}
        \end{fmffile}
        \vspace{0.5cm}
        \caption{Feynman diagrams of the process $e^+ e^- \to W^+ W^-$ in the toy model.}
        \label{fig_toy}
    \end{figure}

We now proceed to computing the amplitude within the framework of the toy model without heavy neutrinos. Let $p_-$ and $p_+$ denote the four-momenta of the incoming electron and positron, and $k_1$ and $k_2$ denote the four-momenta of the outgoing $W^-$ and $W^+$, respectively. In the center-of-mass frame, we define the coordinate system as\footnote{We choose such a coordinate system to simplify expressions for polarization vectors.}

\begin{eqnarray*}
    k_1 & = \mqty(\varepsilon \\ 0 \\ 0 \\ \varepsilon v_W), \hspace{1 cm} k_2 & = \mqty(\varepsilon \\ 0 \\ 0 \\ -\varepsilon v_W), \\
    p_- & = \mqty(\varepsilon \\ \varepsilon \sin\theta \\ 0 \\ \varepsilon \cos\theta), \hspace{1 cm} p_+ & = \mqty(\varepsilon \\ -\varepsilon\sin\theta \\ 0 \\ -\varepsilon\cos\theta), \\
\end{eqnarray*}
\begin{align*}
    e^{(-)}_\mu (+1) &= \frac{1}{\sqrt{2}} \mqty(0 \\ 1 \\ -i \\ 0),  
    &
    e^{(-)}_\mu (-1) &= \frac{1}{\sqrt{2}} \mqty(0 \\ 1 \\ i \\ 0), 
    &
    e_\mu^{(-)}(0) &= \frac{1}{M_W}\mqty(\varepsilon v_W \\ 0 \\ 0 \\ \varepsilon), \\
    e^{(+)}_\mu (+1) &= e^{(-)}_\mu (-1),
    &
    e^{(+)}_\mu (-1) &= e^{(-)}_\mu (+1),
    &
    e_\mu^{(+)}(0) &= \frac{1}{M_W}\mqty(-\varepsilon v_W \\ 0 \\ 0 \\ \varepsilon).
\end{align*}
Here $\theta$ is the angle between $e^-$ and $W^-$, $2\varepsilon = \sqrt{s}$ is the total energy of the system, and \mbox{$v_W = \sqrt{1-\frac{4M_W^2}{s}}$} is the velocity of the $W$-boson; $e^{(-)}$ and $e^{(+)}$ are the polarization four-vectors of $W^-$ and $W^+$, respectively. The amplitude of the process reads:
\begin{eqnarray}
    \mathcal{M}_{fi} = -\frac{ig^2}{4}\cdot\Bigg(\bar{u}(-p_+)\gamma_\mu(1+\gamma_5)u(p_-)\cdot \frac{g^{\mu\nu}-\frac{(p_-+p_+)^\mu(p_-+p_+)^\nu}{M^2}}{s-M^2}\cdot P_{\sigma\rho\nu}e^{(-)\sigma}e^{(+)\rho}+\nonumber \\ +\bar{u}(-p_+)\hat{e}^{(+)}\frac{\hat{p}_--\hat{k}_1}{t}\hat{e}^{(-)}(1+\gamma_5)u(p_-)\Bigg),
\end{eqnarray}
where $P_{\sigma\rho\nu} = g_{\sigma\rho}(k_1-k_2)_\nu + g_{\rho\nu}(2k_2+k_1) - g_{\nu\sigma}(2k_1+k_2)$, $t = \qty(p_--k_1)^2$. After some simplifications, we can write the amplitude in the following form:
\begin{eqnarray}
    \mathcal{M}_{fi} = -\frac{ig^2}{4}\cdot\Bigg(\bar{u}(-p_+)\gamma_\mu(1+\gamma_5)u(p_-)\cdot \frac{1}{s-M^2}\cdot \bigg(\qty(e^{(-)}e^{(+)})(k_1-k_2)^\mu+2e^{(+)\mu}\qty(e^{(-)}k_2)-\nonumber \\-2e^{(-)\mu}\qty(e^{(+)}k_1)\bigg)+\bar{u}(-p_+)\hat{e}^{(+)}\frac{\hat{p}_--\hat{k}_1}{t}\hat{e}^{(-)}(1+\gamma_5)u(p_-)\Bigg).
\end{eqnarray}
Upon squaring the amplitude, summing over the polarizations of $W$-bosons, and substituting into the standard expression for the cross section,
\begin{equation}\label{sech}
    \dd\sigma = \frac{v_W}{32\pi s}|\mathcal{M}_{fi}|^2\dd\cos\theta,
\end{equation}
one obtains the following differential cross section in the limit $\sqrt{s} \gg 2M$:
\begin{equation}\label{SM}
    \frac{\dd\sigma^\text{sim}}{\dd\cos\theta} = \frac{g^4(1+\cos\theta)}{512\pi s(1-\cos\theta)}\cdot \qty(9-2\cos\theta+9\cos^2\theta),
\end{equation}
where ``sim'' stands for the simplest model under consideration. As is evident from the result, the cross section scales with energy as $\frac{1}{s}$. However, when considering the $s$- and $t$-channel contributions separately,\footnote{The result for a separate diagram is not gauge invariant. However, we can perform such a calculation in the unitary gauge.} one finds leading terms that grow with energy as $s$:
\begin{equation}
    \frac{\dd\sigma_s^\text{sim}}{\dd\cos\theta} = \frac{g^4s}{512\pi M^4}\cdot\qty(1-\cos^2\theta),
\end{equation}
\begin{equation}\label{asymp}
    \frac{\dd\sigma_t^\text{sim}}{\dd\cos\theta} = \frac{g^4s}{512\pi M^4}\cdot\qty(1-\cos^2\theta).
\end{equation}
The cancellation of these energy-growing terms arises because the $s$ and $t$-channel amplitudes enter the total amplitude with opposite signs, resulting in their mutual subtraction~\cite{Vainshtein:1971ip,Weinberg:1971fb,Bogomolnyi:1973nj,Okun:ITEPWS1,LQ}. Consequently, the unitarity-consistent asymptotic behavior $\sigma \propto \frac{1}{s}$ is restored. 

Next, we extend the model by introducing a single heavy neutrino (the generalization to multiple states being straightforward), characterized by a mixing parameter $V_{eN}$ and a mass $M_N$. At the end of Sec. \ref{sec:model} it was indicated that we can do it in two ways: exact and approximate. So, in the next two subsections, we derive the cross section separately for these two ways.

\subsection{Exact unitary mixing}
\label{sec:exact}

The total amplitude remains formally identical to that without the heavy neutrino, except that the electron neutrino propagator is replaced by the linear combination of the propagators for the light and heavy neutrino states ($M_N$ in the numerator of the second term does not contribute due to the projector $\frac{1+\gamma_5}{2}$):
\begin{align}\label{comb_prop}
    (1-|V_{eN}|^2)\frac{\hat{p}_--\hat{k}_1}{M^2-2(p_-k_1)}+ & |V_{eN}|^2\frac{\hat{p}_--\hat{k}_1}{M^2-2(p_-k_1)-M_N^2} = \nonumber \\  & = \frac{\hat{p}_--\hat{k}_1}{M^2-2(p_-k_1)}
    \cdot\frac{M^2-2(p_-k_1)-M_N^2(1-|V_{eN}|^2)}{M^2-2(p_-k_1)-M_N^2}.
\end{align}
The modification of the light-neutrino coupling is taken into account to ensure that the introduction of the heavy neutrino preserves unitarity. Performing a calculation analogous to the one above yields (``un'' stands for unitary mixing) 
\begin{eqnarray}\label{heavy_acc}
    \frac{\dd\sigma^\text{un}}{\dd\cos\theta} = \frac{g^4|V_{eN}|^4M_N^4s(1-\cos^2\theta)}{128\pi M^4(s+2M_N^2-s\cos\theta)^2},
\end{eqnarray}
valid in the limit $\sqrt{s} \gtrsim \frac{M\sqrt{3}}{|V_{eN}|}$. This expression can be used to study the behavior of the differential cross section in the limits $\sqrt{s} \ll M_N$ and $\sqrt{s} \gg M_N$, which will be discussed in Sec.~\ref{See_Saw_exten}.

We now proceed to analyze the key features that arise in the case of an exact implementation of heavy neutrino states. Before proceeding, it is important to note that we will restrict our analysis to the central angular region. This choice is motivated by the fact that the total (integrated) cross section is dominated by the pronounced peak near $\cos\theta=1$, whereas the contribution from heavy neutrino exchange becomes substantial in the region $\cos\theta \sim 0$ ($|t| \approx \frac{s}{2}$), which corresponds precisely to the central part of the angular distribution.

\begin{enumerate}
    \item Let us compare formulas (\ref{SM}) and (\ref{heavy_acc}) in the region $\sqrt{s} \ll M_N$ at $\cos\theta=0$. From these expressions, it follows that the cross section is significantly modified if the following condition is satisfied:
    \begin{equation}\label{s_cond1}
        \sqrt{s} \gtrsim \frac{M\sqrt{3}}{|V_{eN}|}.
    \end{equation}
    It is also clear that a sizable correction can occur only if the heavy neutrino mass fulfills the relation: 
    \begin{equation}\label{first_cond}
        M_N \gtrsim \frac{M\sqrt{3}}{|V_{eN}|}.
    \end{equation}
    What it really means is that if the heavy neutrino mass can be neglected in the numerator of \eqref{comb_prop}, i.e., $\abs{V_{eN}}^2 M_N^2\ll M^2$, then we would just get the massless propagator and the correction is negligible (see the line for $M_N = 1$ TeV in Fig.~\ref{toy_model}, for these parameters $M_N\approx M\sqrt{3}/\abs{V_{eN}}$). 
    \item When $\sqrt{s} \ll M_N$ and the condition (\ref{first_cond}) holds, one can neglect the terms proportional to $s$ in the denominator of (\ref{heavy_acc}), resulting in a linear dependence on $s$:
    \begin{equation}\label{lin_grow}
        \qty(\frac{\dd\sigma^\text{un}}{\dd\cos\theta})_{\sqrt{s} \ll M_N} = \frac{g^4|V_{eN}|^4s}{512\pi M^4}\cdot\qty(1-\cos^2\theta).
    \end{equation}
    Such a growth is expected to occur within the range 
    \begin{equation}\label{region_1}
        \frac{M\sqrt{3}}{|V_{eN}|} \lesssim \sqrt{s} \lesssim M_N.
    \end{equation}
    From this expression, one can see that the correction induced by the introduction of a heavy neutrino increases with the mixing parameter $|V_{eN}|^2$, but does not depend on the heavy neutrino mass $M_N$.
    \item Next, let us consider (\ref{heavy_acc}) in the high-energy limit $\sqrt{s} \gg M_N$. Neglecting $M_N^2$ in the denominator, we obtain the following result: 
    \begin{equation}\label{exact_decrease}
        \qty(\frac{\dd\sigma^\text{un}}{\dd\cos\theta})_{\sqrt{s} \gg M_N} = \frac{g^4|V_{eN}|^4M_N^4\qty(1-\cos^2\theta)}{128\pi M^4s\qty(1-\cos\theta)^2}.
    \end{equation}
    This expression shows that, for energies above the heavy neutrino mass, the cross section decreases as $\frac{1}{s}$, in agreement with the unitarity constraint on the $S$-matrix.
\end{enumerate}

As will be demonstrated in the following section, the behavior of the cross section in the case of linearized mixing differs significantly from the results obtained in this section.

From the obtained formula, one can see that, at asymptotically high energies (when $\sqrt{s} \gg M_N$), the cross section again approaches the unitarity limit $\frac{1}{s}$. This occurs due to the cancellation between the $s$- and $t$-channel amplitudes. To make this cancellation explicit, let us consider $A_s$, $A_t$, and $A_N$, which are the amplitudes of the $s$-channel, $t$-channel, and heavy neutrino, respectively. Under the assumption of unitary mixing between light and heavy neutrino states, we obtain
\begin{equation}\label{ampl_1}
    A = A_s+\qty(1-|V_{eN}|^2)A_t+|V_{eN}|^2A_N = A_s+A_t+|V_{eN}|^2(A_N-A_t).
\end{equation}
The sum of the first two amplitudes remains constant at large $s$ [see~\eqref{SM}--\eqref{asymp}], while the behavior of $A_N - A_t$ is determined by the difference between the propagators:
\begin{equation}
    \frac{\hat{p}_--\hat{k}_1}{(p_--k_1)^2-M_N^2}-\frac{\hat{p}_--\hat{k}_1}{(p_--k_1)^2} = \frac{\qty(\hat{p}_--\hat{k}_1)M_N^2}{(p_--k_1)^2\qty((p_--k_1)^2-M_N^2)}.
\end{equation}
Thus, although each individual amplitude, $A_t$ and $A_N$, grows proportionally to $s$ [see~\eqref{asymp}], their difference remains constant, leading to the $\frac{1}{s}$ dependence of the cross section. 

\subsection{Linearized mixing}

We now turn to the case of a linearized implementation of heavy-light neutrino mixing. In this scheme, the sum of propagators differs from that obtained in Sec.~\ref{sec:exact} and takes the form:
\begin{align}
    \frac{\hat{p}_--\hat{k}_1}{M^2-2(p_-k_1)}+ & |V_{eN}|^2\frac{\hat{p}_--\hat{k}_1}{M^2-2(p_-k_1)-M_N^2} = \nonumber \\ 
    & = \frac{\hat{p}_--\hat{k}_1}{M^2-2(p_-k_1)}
    \cdot\frac{\qty(1+|V_{eN}|^2)\qty(M^2-2(p_-k_1))-M_N^2}{M^2-2(p_-k_1)-M_N^2}.
\end{align}
As a result, the differential cross section is given by (``lin'' stands for linearized mixing)
\begin{equation}\label{heavy_lin}
    \frac{\dd\sigma^\text{lin}}{\dd\cos\theta} = \frac{g^4|V_{eN}|^4s^3(1-\cos^2\theta)(1-\cos\theta)^2}{512\pi M^4(s+2M_N^2-s\cos\theta)^2},
\end{equation}
which is valid in the limit $\sqrt{s} \gtrsim \max\qty(\sqrt{\frac{MM_N}{|V_{eN}|}}, \frac{M\sqrt{3}}{|V_{eN}|})$. In close analogy with Sec.~\ref{sec:exact}, we now discuss the main features that arise in the presence of linearized mixing.
\begin{enumerate}
    \item We compare (\ref{SM}) and (\ref{heavy_lin}) in the regime $\sqrt{s} \ll M_N$ at $\cos\theta=0$. Neglecting the terms proportional to $s$ in the denominator of (\ref{heavy_lin}), we obtain
    \begin{equation}\label{lin_low}
        \qty(\frac{\dd\sigma^\text{lin}}{\dd\cos\theta})_{\sqrt{s} \ll M_N} = \frac{g^4|V_{eN}|^4s^3}{2048\pi M^4M_N^4}.
    \end{equation}
    By comparing this expression with (\ref{SM}), we find that the contribution of the heavy neutrino becomes comparable to the case without a heavy neutrino when the following condition is satisfied [please note the validity domain of~\eqref{heavy_lin}]
    \begin{equation}\label{s_cond2}
        \sqrt{s} \gtrsim \max\qty(\sqrt{\frac{MM_N}{|V_{eN}|}}, \frac{M\sqrt{3}}{|V_{eN}|}).
    \end{equation}
    In contrast to the result in Sec.~\ref{sec:exact}, we would \emph{always} get a sizable correction. However, if
    \begin{equation}\label{second_cond}
        M_N \gtrsim \frac{M}{|V_{eN}|},
    \end{equation}
    the cross section has a more complicated behavior (see the next item). 
    \item If $\sqrt{s} \ll M_N$ and the condition (\ref{second_cond}) holds, then according to (\ref{lin_low}), the cross section exhibits cubic growth, $\sigma\propto s^3$, within the range:
    \begin{equation}\label{region_2}
        \sqrt{\frac{MM_N}{|V_{eN}|}} \lesssim \sqrt{s} \lesssim M_N.
    \end{equation}
    Here, the size of the correction is controlled by the parameters $\frac{M}{|V_{eN}|}$ and $M_N$, and it increases with larger values of $|V_{eN}|$ and decreases with larger values of $M_N$.
    \item We now consider the opposite regime $\sqrt{s} \gg M_N$. Neglecting the heavy neutrino mass in (\ref{heavy_lin}), we obtain the following expression:
    \begin{equation}\label{non-unit_grow}
        \qty(\frac{\dd\sigma^\text{lin}}{\dd\cos\theta})_{\sqrt{s} \gg M_N} = \frac{g^4|V_{eN}|^4s}{512\pi M^4}.
    \end{equation}
    In contrast to the case of exact unitary mixing, at high energies, the cross section now exhibits unbounded linear growth, $\sigma\propto s$, in clear conflict with the $S$-matrix unitarity. Let us note that there is no dependence on $M_N$ in~\eqref{non-unit_grow}.
\end{enumerate}

Let us demonstrate at the amplitude level why there is no cancellation in this case. 
If the heavy neutrino is simply added without adjusting the coupling of the light neutrino, the resulting expression
\begin{equation}\label{ampl_2}
    A = A_s+A_t+|V_{eN}|^2A_N,
\end{equation}
in addition to the sum $A_s+A_t$, which behaves like a constant at large $s$, contains a term $|V_{eN}|^2A_N$ that grows with energy proportionally to $s$, thereby violating the unitarity bound on the cross section. It is therefore crucial that heavy neutrino states be incorporated in a unitary manner to preserve the asymptotic behavior $\sigma \propto \frac{1}{s}$ at high energies.

\subsection{Comparison of two approaches}

\begin{figure}[t]
    \centering
    \includegraphics[width=6in]{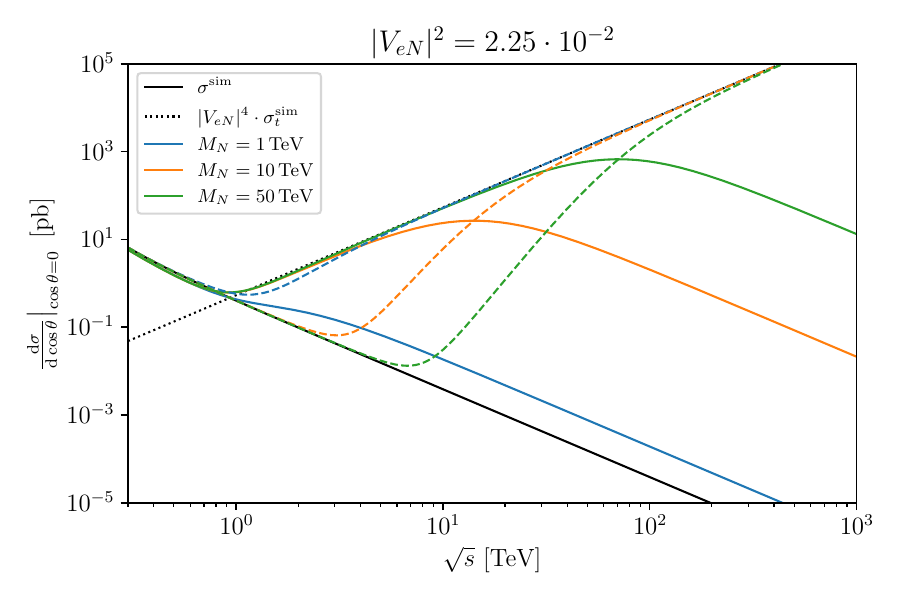}
    \caption{Dependence of differential cross section at $\cos\theta = 0$ on the value of $\sqrt{s}$ for different masses $M_N$. Here, solid lines correspond to exactly unitary mixing, while dashed lines correspond to linearized mixing.}
    \label{toy_model}
\end{figure}

In conclusion of this section, we compare the results obtained from the exact unitary and linearized implementations of heavy-light neutrino mixing; see Fig.~\ref{toy_model}. Here, we consider the mixing parameter $|V_{eN}|^2 = 2.25\cdot 10^{-2}$ since a larger mixing strength enhances the effects under study.

First, we note that the sizable corrections due to heavy neutrinos appear at different energies in these two cases. If we look at the lines for $M_N =1$ TeV, we can see that in exactly unitary mixing there is only a slight modification in the cross section (in comparison to other lines), while in the linearized case the cross section almost immediately demonstrates its asymptotic behavior. Such behavior arises from (\ref{first_cond}) in the exactly unitary case and from (\ref{s_cond2}) in the linearized case. However, when we increase the heavy neutrino mass to $M_N =10$ TeV (see orange lines), the cross sections in both cases change their behavior. In exactly unitary mixing, the cross section starts to grow linearly with $s$ [see (\ref{lin_grow})] until the energy becomes comparable to the heavy neutrino mass $\sqrt{s} \sim M_N$, and decreases as $1/s$ when $\sqrt{s} \gg M_N$ [see~(\ref{exact_decrease})]. In linearized mixing, the cross section is not modified until $\sqrt{s} \sim \sqrt{\frac{MM_N}{|V_{eN}|}}$ [see (\ref{s_cond2})], then it starts to grow cubically, $\sigma\propto s^3$ [formula~(\ref{lin_low})]. When $\sqrt{s} \gtrsim M_N$, the cross section approaches linear behavior, $\sigma\propto s$ [see (\ref{non-unit_grow})]. If we continue to increase the heavy neutrino mass up to $M_N =50$ TeV, the overall behavior of the cross sections retains their features (see the green lines), just making the modifications more significant. It is important to mention that modifications in both cases become comparable around the heavy neutrino mass scale.

It is worth noting that in the region defined by (\ref{region_1}), exact unitary mixing exhibits the same linear asymptotic behavior, given by $|V_{eN}|^4\cdot\sigma^\text{sim}_t$, as the linearized scenario does in the $\sqrt{s} \gg M_N$ regime (see Fig.~\ref{toy_model}). The origin of this similarity can be traced to the fact that, in the exact unitary case with $\sqrt{s} \ll M_N$, the heavy neutrino contribution to the amplitude can be neglected, so that the term $|V_{eN}|^2A_t$ in (\ref{ampl_1}) is effectively subtracted from the amplitude $A_t$. By contrast, in the linearized case at $\sqrt{s} \gg M_N$, the same quantity $|V_{eN}|^2A_t$ is \textit{added} to $A_t$. Therefore, the leading term in the cross section is 
\begin{equation}
\frac{\abs{\pm \abs{V_{eN}}^2 A_t}^2}{32\pi s}=\abs{V_{eN}}^4\frac{\dd\sigma^\text{sim}_t}{\dd\cos\theta}. 
\end{equation}
The difference in sign manifests itself only in the interference term, which, at high energies, can be neglected on a logarithmic scale.

If we consider a very heavy neutrino $M_N \to \infty$, we can see from Fig.~\ref{toy_model} that linearized mixing will not modify the cross section [see (\ref{lin_low})]. However, in the exact unitary mixing, the modification of the cross section depends solely on the mixing parameter $|V_{eN}|^2$ and does not depend on the heavy neutrino mass $M_N$ [see (\ref{lin_grow})]. This makes the exact unitary framework particularly well suited for deriving experimental constraints on the mixing parameter $|V_{eN}|^2$ for arbitrarily large masses of heavy neutrinos. According to (\ref{s_cond1}), even a relatively rough measurement would already allow one to set a bound of the form
\begin{equation}
    |V_{eN}|^2 \lesssim \frac{3M^2}{s}
\end{equation}
on the mixing parameter. If, in addition, a high-precision measurement is feasible, the constraint on $|V_{eN}|^2$ can be improved using data at lower center-of-mass energies (see the region before the intersection of $\sigma^\text{sim}$ and $|V_{eN}|^4\cdot\sigma_t^\text{sim}$), where the cross section is still noticeably modified. A more detailed analysis of the dependence of the differential cross section on the mixing parameter will be presented in Sec.~\ref{See_Saw_exten}.

\section{$e^+ e^- \to W^+ W^-$ process in the Standard Model seesaw extension}
\label{See_Saw_exten}

We now proceed to compare the toy model with the seesaw type-I scenario. In the SM seesaw type-I extension with $\mathcal{N}$ heavy neutrinos, the process $e^+e^-\to W^+W^-$ is represented by $\mathcal{N}+6$ Feynman diagrams: three corresponding to the $s$-channel and $\mathcal{N}+3$ to the $t$-channel (see Fig.~\ref{fig1}). 
    \begin{figure}[t]
    \centering
        \begin{fmffile}{cross2}
            \begin{fmfgraph*}(180,120)
                \fmfleft{i1,i2}
                \fmfright{o1,o2}
                \fmf{fermion}{v1,i2}
                \fmf{fermion}{i1,v1}
                \fmf{boson, label=$\gamma,, Z,, H$}{v1,v2}
                \fmf{boson}{v2,o2}
                \fmf{boson}{v2,o1}
                \fmflabel{$e^+$}{i2}
                \fmflabel{$e^-$}{i1}
                \fmflabel{$W^-$}{o1}
                \fmflabel{$W^+$}{o2}
            \end{fmfgraph*}
        \end{fmffile}
        \hspace{0.7 cm}
        \begin{fmffile}{cross3}
            \begin{fmfgraph*}(180,120)  
                \fmfleft{i1,i2}
                \fmfright{o1,o2}
                \fmf{photon}{v2,o2}
                \fmf{photon}{o1,v1}
                \fmf{fermion, label=$\nu_i,, N_I$}{v1,v2}
                \fmf{fermion}{i1,v1}
                \fmf{fermion}{v2,i2}
                \fmflabel{$e^+$}{i2}
                \fmflabel{$e^-$}{i1}
                \fmflabel{$W^-$}{o1}
                \fmflabel{$W^+$}{o2}
            \end{fmfgraph*}
        \end{fmffile}
        \vspace{0.5cm}
        \caption{Feynman diagrams of the process $e^+ e^- \to W^+ W^-$.}
        \label{fig1}
    \end{figure}

\begin{figure}[t]
    \centering
    \includegraphics[width=6in]{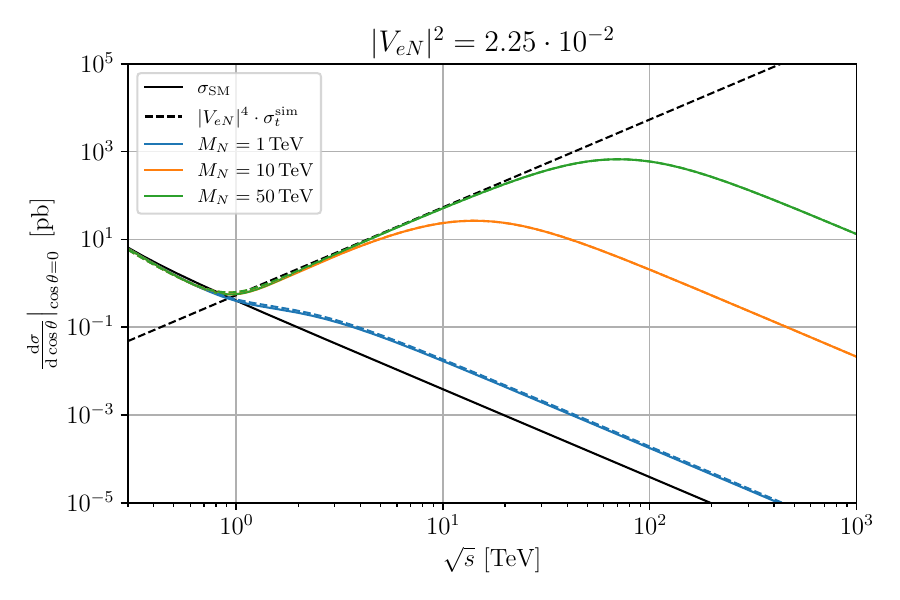}
    \caption{Dependence of differential cross section at $\cos\theta = 0$ on the value of $\sqrt{s}$ for different masses $M_N$. Here, solid lines correspond to the SM extension, while dashed lines correspond to the toy model.}
    \label{SM_toy}
\end{figure}

To begin with, we compare the toy model and the SM seesaw type-I extension in the case of $\mathcal{N}=1$. In Fig.~\ref{SM_toy}, we show the differential cross sections at $\cos\theta=0$ for the Standard Model extended with a single heavy neutrino (solid curves) and for the toy model with one heavy neutrino (dashed curves). As seen from this figure, the toy model curves do not coincide exactly with the Standard Model ones due to the effect of electroweak $\gamma-Z$ mixing in the latter. When $M_N = 1$~TeV, the difference is approximately 10\% at any energies. This happens because $M_N=1$~TeV almost violates the condition~(\ref{first_cond}), which means that the SM part of the cross section is not negligible. Therefore, it is clear that contributions from all channels should be taken into account as  accurately as possible. That is why the toy model is not so exact here. 

However, when $M_N = 10$~or~$50$~TeV, the same difference occurs only for energies $\sqrt{s}\lesssim 1$~TeV, and rapidly decreases at higher energies. This occurs because $M_N$ values now satisfy the condition (\ref{first_cond}), and the leading term (\ref{exact_decrease}) is defined by heavy neutrino contribution according to~\eqref{ampl_1} and does not depend on $\gamma-Z$ mixing.

Summing up, the toy model reproduces all qualitative features of the Standard Model. Moreover, the toy model has more accuracy when the difference with the SM is increasing. Therefore, the conclusions of the previous section remain valid in the full theory.

We now turn to the case where $\mathcal{N}=3$ neutrinos are introduced, partially discussed in Sec.~\ref{sec:model}. In the process under consideration, we impose $m_{\nu_1} = m$, $m_{\nu_2} = m_{\nu_3} = 0$, $|V_{e1}|^2 = \frac{M_1}{M_2}|V_{eN}|^2$, \mbox{$|V_{e2}|^2 \equiv |V_{eN}|^2$}, $|V_{e3}|^2 = \frac{M_2}{M_3}|V_{eN}|^2$ in the mass hierarchy $M_1 \ll M_2 \ll M_3$ [see \cite{Asaka:2015oia}]. In the following plots, we take $\frac{M_1}{M_2} = \frac{M_2}{M_3}=10^{-4}$ to clearly outline the main features of the case. We take $m_{\nu_1}$ as nonzero to satisfy the seesaw relation (\ref{seesaw}) for our mixing parameters. These parameter choices are implemented in our \texttt{FeynRules} model \texttt{HeavyNU},\footnote{It is called \texttt{HeavyNU} since it is based on \texttt{HeavyN}, and ``\texttt{U}'' stands for unitary neutrino mixing.} which realizes exact unitary mixing with three heavy neutrinos. Let us note that \texttt{HeavyNU} is easily adoptable to many scenarios as far as the full neutrino mixing matrix is unitary. We attach model files to our work~\cite{Chachava:2026mbu}.  Numerical calculations are performed with the help of this model; however, let us present some analytical results as well.

\begin{figure}[t]
    \centering
    \includegraphics[width=6in]{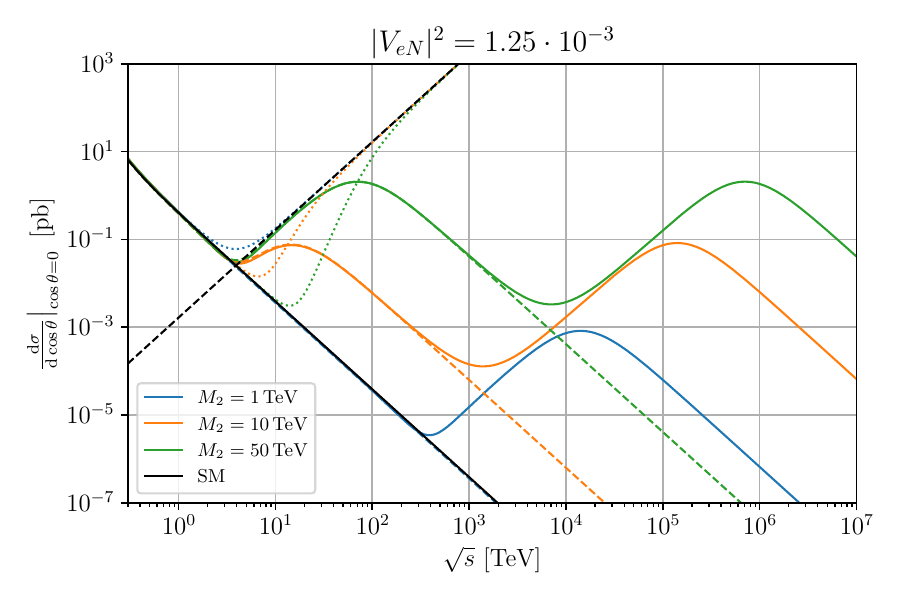}
    \caption{Dependence of differential cross section at $\cos\theta = 0$ on the value of $\sqrt{s}$ for different masses $M_2$. Here, solid lines correspond to SM extension with three heavy neutrinos, dashed lines correspond to toy model with one heavy neutrino, and dotted lines correspond to \texttt{HeavyN} model with three heavy neutrinos.}
    \label{3HNL}
\end{figure}

The sum of the amplitudes corresponding to the heavy neutrinos is as follows: 
\begin{equation}\label{See_Saw_ampl}
    \mathcal{M}_{fi}^\mathcal{N} = -\frac{g^2}{2}\bar{u}(-p_+)\hat{e}^{(+)}(b)\frac{1+\gamma_5}{2}\Bigg[|V_{e1}|^2\frac{\hat{p}_--\hat{k}_1}{t-M_1^2}+|V_{e2}|^2\frac{\hat{p}_--\hat{k}_1}{t-M_2^2}+|V_{e3}|^2\frac{\hat{p}_--\hat{k}_1}{t-M_3^2}\Bigg]\hat{e}^{(-)*}(a)\frac{1+\gamma_5}{2}u(p_-),
\end{equation}
where $a$ and $b$ denote polarizations of $W^-$ and $W^+$ respectively.
Using the discussed mass hierarchy and mixing parameters, the sum in the square brackets can be simplified as
\begin{align}\label{sum_3}
    |V_{e1}|^2\cdot\frac{\hat{p}_--\hat{k}_1}{t-M_1^2}+|V_{e2}|^2\cdot\frac{\hat{p}_--\hat{k}_1}{t-M_2^2}+|V_{e3}|^2\cdot\frac{\hat{p}_--\hat{k}_1}{t-M_3^2} \approx \frac{\qty(t-M_1M_2)|V_{eN}|^2}{t\qty(t-M_2^2)}\qty(\hat{p}_--\hat{k}_1)
\end{align}
in the limit $M_1 \ll \sqrt{s} \ll M_3$. As discussed in Sec.~\ref{Toy_Model_sec}, we focus on the differential cross section near $\cos\theta =0$. At this point, $|t|=\frac{s}{2}$, and when $s \ll M_1M_2$, we find that the linear combination of the propagators is proportional to $\frac{M_1}{M_2} \ll 1$ due to the mass hierarchy, so it can be neglected. From the expression above, it follows that the presence of two additional heavy neutrinos has a negligible impact on the cross section while $\sqrt{s} \ll M_3$, and we reproduce the results of the case with one heavy neutrino $M_N = M_2$. However, at higher energies, the cross section starts to increase again due to the contribution proportional to $|V_{e3}|^2$, up to $\sqrt{s} \sim M_3$, and then decreases once more, approaching the asymptotic regime 
\begin{equation}\label{second_decrease}
    \qty(\frac{\dd\sigma^\mathcal{N}}{\dd\cos\theta})_{\sqrt{s} \gg M_3} = \frac{g^4|V_{e3}|^4M_3^4\qty(1-\cos^2\theta)}{128\pi M_W^4s\qty(1-\cos\theta)^2} = \frac{g^4|V_{eN}|^4M_2^2M_3^2\qty(1-\cos^2\theta)}{128\pi M_W^4s\qty(1-\cos\theta)^2},
\end{equation}
see (\ref{exact_decrease}). The characteristic transitions between these regimes are illustrated in Fig.~\ref{3HNL}.

We note that the transition associated with the scale $M_3$ occurs at very high energies and therefore most likely cannot be probed experimentally. For realistic collider energies, one can safely use the expression (\ref{sum_3}). However, it is interesting from the theoretical point of view.

In contrast, for linearized mixing, no such transitions appear for the following reason. As can be seen from (\ref{non-unit_grow}), the asymptotic behavior in the limit $\sqrt{s} \gg M_N$ is independent of the heavy neutrino mass and depends only on its mixing parameter. Hence, the neutrino with mass $M_3$ contributes at the level of $|V_{e3}|^2$, which is smaller by a factor of $10^4$ than the contribution from the second neutrino with mixing $|V_{eN}|^2$, and thus its effect on the cross section is negligibly small.

\begin{figure}[p]
    \centering
    \begin{minipage}{1.0\textwidth}
        \centering
        \includegraphics[width=6in]{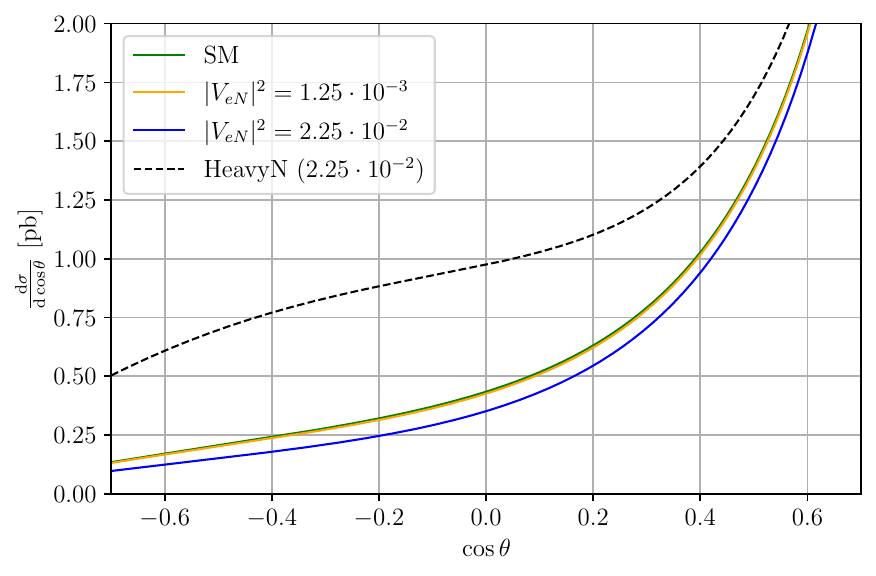}
        \caption{Angular distributions at $\sqrt{s} = 1$ TeV and $M_N = 0.5$ TeV.}
        \label{fig3}
    \end{minipage}
    \vspace{0.02\textwidth}
    \begin{minipage}{1.0\textwidth}
        \centering
        \includegraphics[width=6in]{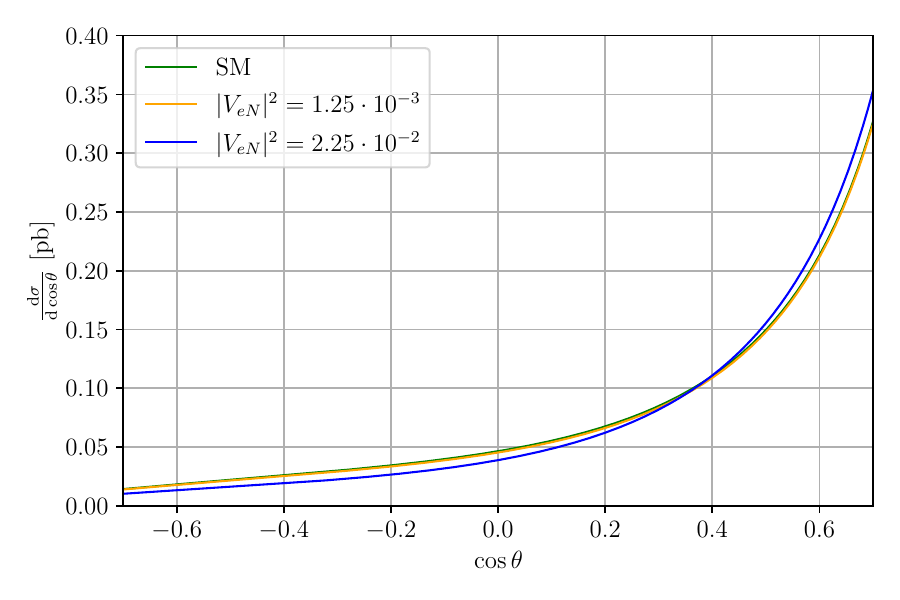}
        \caption{Angular distributions at $\sqrt{s} = 3$ TeV and $M_N = 0.5$ TeV.}
        \label{fig4}
    \end{minipage}
\end{figure}

\begin{figure}[p]
    \centering
    \begin{minipage}{1.0\textwidth}
        \centering
        \includegraphics[width=6in]{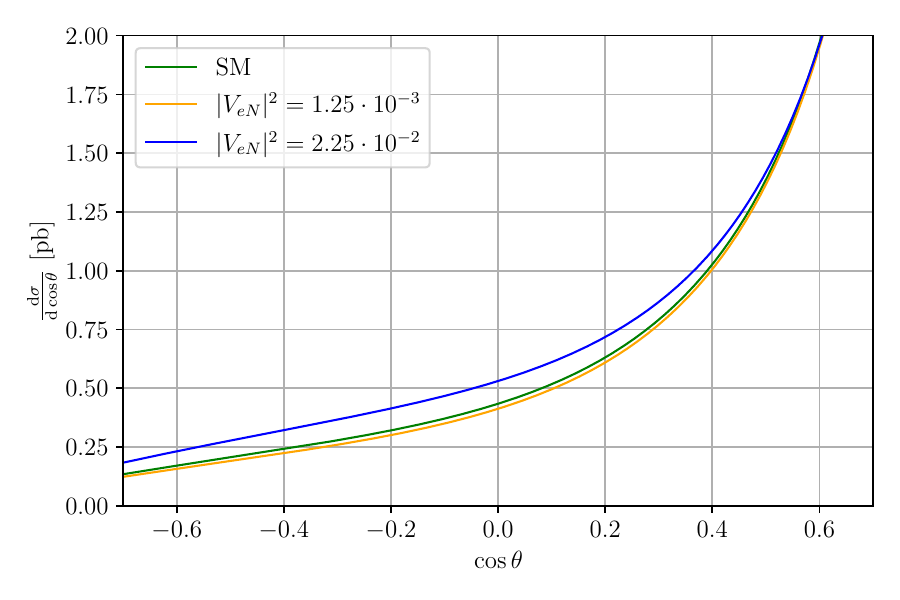}
        \caption{Angular distributions at $\sqrt{s} = 1$ TeV and $M_N = 3$ TeV.}
        \label{fig6}
    \end{minipage}
    \vspace{0.02\textwidth}
    \begin{minipage}{1.0\textwidth}
        \centering
        \includegraphics[width=6in]{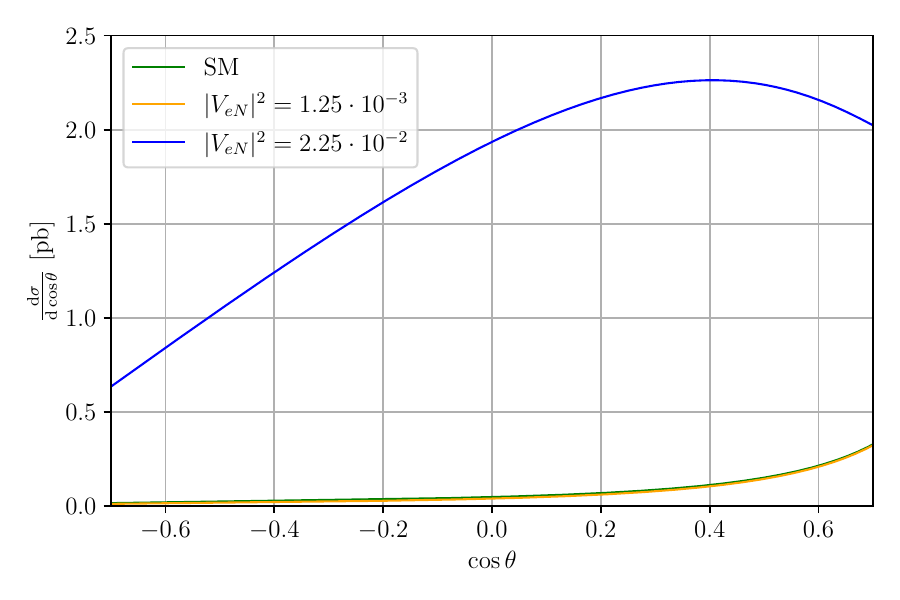}
        \caption{Angular distributions at $\sqrt{s} = 3$ TeV and $M_N = 3$ TeV.}
        \label{fig7}
    \end{minipage}
\end{figure}

The simulation of the $e^+e^- \to W^+W^-$ process was performed using \texttt{MadGraph}, and the obtained results are summarized below. Let us begin with the analysis of the plots shown in Fig.~\ref{fig3}–\ref{fig7}, which depict the dependence of the differential cross section on the scattering angle. The analysis of these results allows us to draw several conclusions.

Figure~\ref{fig3} should be compared with Fig.~6 from~\cite{Drutskoy:2024hqx}. Let us note that in~\cite{Drutskoy:2024hqx} the reconstruction of all events was performed and the sign of $W$ was not reconstructed, so we can compare only the shape of the symmetrized ($\cos\theta\Leftrightarrow -\cos\theta$) distributions. The \texttt{HeavyN} line agrees between these two plots, predicting doubled number of events in comparison to the SM for heavy neutrinos with these parameters. At the same time, \texttt{HeavyNU} predicts the diminishment of the cross section.

As illustrated in Fig.~\ref{fig6}, the cross section remains below the Standard Model prediction for small mixing parameter but exceeds it for larger mixing parameter. This behavior can be understood as follows. As demonstrated in \cite{Yamatsu:2023bde}, the $t$-channel diagrams contribute more significantly to the total cross section than the $s$-channel diagrams. As it was explained in Sec.~\ref{Toy_Model_sec}, an increase in the heavy neutrino mass suppresses the $t$-channel contribution through the $(1-|V_{eN}|^2)$ factor. This enhances the relative importance of the $s$-channel, which causes the total cross section to rise with energy. However, before the $s$-channel begins to dominate, its magnitude becomes comparable to that of the $t$-channel. Because these two contributions interfere destructively when both $W$-bosons are longitudinally polarized, the total cross section temporarily decreases. However, the minimal value of the cross section is non-zero due to the contribution of the transverse polarizations of the $W$-bosons. For example, if $W$-bosons have transverse polarizations, then diagrams with $\gamma$ and $Z$ do not contribute (it can be seen from the structure of $\gamma WW$ and $ZWW$ vertices). This means that only diagrams with the Higgs boson and neutrinos contribute to the cross section, and even when the longitudinal part of the cross section becomes zero, the total cross section remains positive. At higher energies, the $s$-channel eventually surpasses the $t$-channel, leading to a substantial enhancement of the cross section. Naturally, the resulting growth also depends on the magnitude of the mixing parameter $|V_{eN}|^2$.

For a heavy neutrino mass $M_N = 500$ GeV, the correction due to the presence of heavy neutrinos is negligible across all considered center-of-mass energies and mixing parameters. In contrast, when the heavy neutrino mass is set to $M_N = 3$ TeV, the cross section could be significantly modified. For $\sqrt{s} = 3$~TeV and $|V_{eN}|^2 = 2.25\cdot 10^{-2}$, it becomes several times larger than the Standard Model cross section (see Fig.~\ref{fig7}). In order to understand this, let us go back to the region~\eqref{region_1} defining where the cross section grows with energy (while the SM cross section is always decreasing as $1/s$ for the considered energies). For $M_N=500$~GeV and all considered mixing parameters, this region is empty, so we did not observe any enhancement. The same is true for $M_N=3$~TeV and $\abs{V_{eN}}^2=1.25\cdot 10^{-3}$. However, for $M_N=3$~TeV and $\abs{V_{eN}}^2=2.25\cdot 10^{-2}$, there is a region from 1 to 3 TeV where the cross section grows as $s$. This explains why only this line gets enhanced in Fig.~\ref{fig7}.

\begin{figure}[p]
    \centering
    \begin{minipage}{1.0\textwidth}
        \centering
        \includegraphics[width=6in]{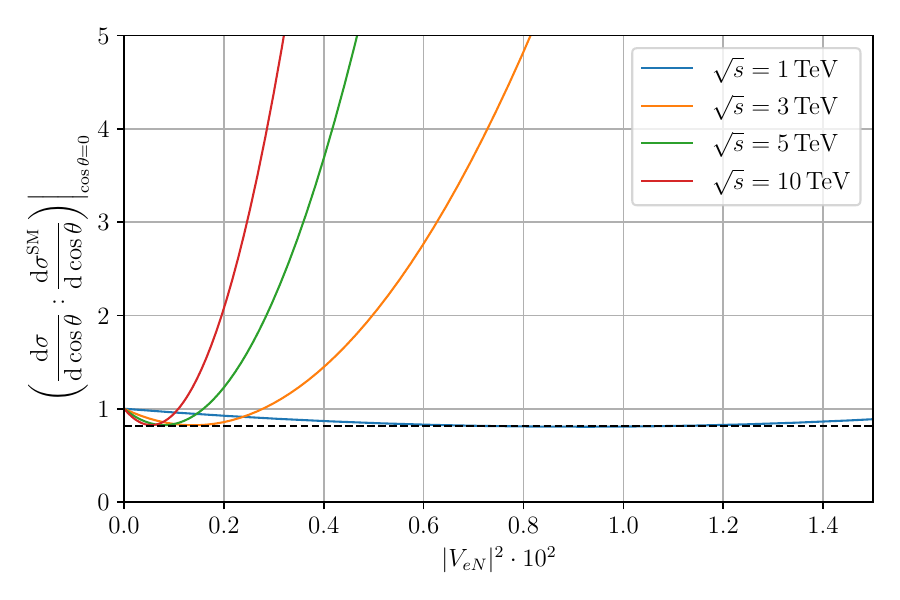}
        \caption{Dependence of the cross section on the value of $|V_{eN}|^2$ for heavy neutrino mass $M_N = 3$~TeV.}
        \label{fig9}
    \end{minipage}
    \vspace{0.02\textwidth}
    \begin{minipage}{1.0\textwidth}
        \centering
        \includegraphics[width=6in]{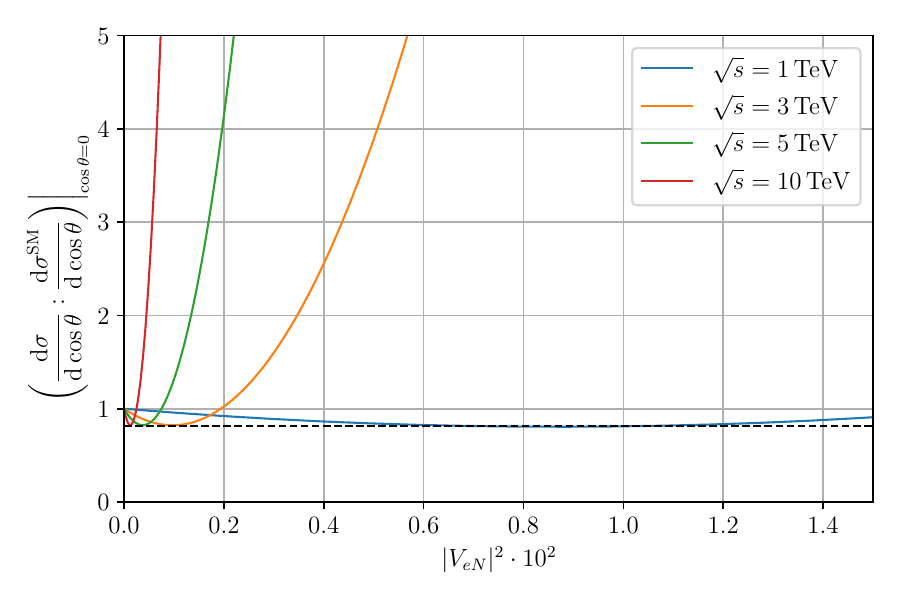}
        \caption{Dependence of the cross section on the value of $|V_{eN}|^2$ for heavy neutrino mass \mbox{$M_N = 10$}~TeV.}
        \label{fig10}
    \end{minipage}
\end{figure}

Figures~\ref{fig9} and \ref{fig10} show how the ratio of the modified cross section to the SM cross section depends on the mixing parameter for different energies at $\cos\theta=0$. A similar effect is clearly visible in these plots: for small $|V_{eN}|^2$, the $s$-channel becomes comparable to the $t$-channel, leading to a decrease in the cross section, while for large $|V_{eN}|^2$, the $s$-channel becomes dominant, resulting in a noticeable increase in the cross section. As the collision energy increases, the suppression of the $t$-channel occurs more rapidly, and for large $|V_{eN}|^2$, a significant enhancement of the total cross section is observed. Let us also mention that the differential cross section has a minimal value. The location of the minimum can be found from the analysis of the accurate formula for the cross section, which is a quadratic polynomial of the argument $|V_{eN}|^2$; see~\eqref{See_Saw_ampl} and~\eqref{sum_3}. In the toy model, the cross section reaches its minimal value when the mixing parameter $|V_{eN}|^2$ becomes equal to
\begin{equation}
   \label{eq:V_min}
    |V_{eN}|^2_\text{min} = \frac{M_W^2\qty(s+2M_N^2)}{2M_N^2s}
\end{equation}
in the limit $\sqrt{s}\gg M_W$. The minimal value of the cross section is
\begin{equation}
   \label{eq:minimum}
    \left.\qty(\dfrac{\mathrm{d}\sigma^\text{un}}{\mathrm{d}\cos\theta}:\dfrac{\mathrm{d}\sigma^\text{sim}}{\mathrm{d}\cos\theta})\right|_{\cos\theta=0} = \frac{8}{9}.
\end{equation}
This value does not depend on the center-of-mass energy $\sqrt{s}$ and heavy neutrino mass $M_N$ when $\sqrt{s}\gg M$. In Figs.~\ref{fig9} and \ref{fig10}, however, the minimal values of the cross sections are lower than~$\frac{8}{9}$. As we have mentioned before, this happens due to the $\gamma$---$Z$ mixing in the Standard Model, which does not occur in the toy model. The minimal value of the cross section in this Standard Model extension can be found using \texttt{FeynCalc}~\cite{Shtabovenko:2023idz}:
\begin{align}\label{eq:newmin}
    \qty(\abs{V_{eN}}^2)_\text{min}^\mathcal{N} & = \frac{M_W^2\qty(s+2M_N^2)}{2M_N^2s\cdot \cos^2\theta_W},\\
    \label{eq:seesaw_min}
    \left.\qty(\dfrac{\mathrm{d}\sigma^\mathcal{N}}{\mathrm{d}\cos\theta}:\dfrac{\mathrm{d}\sigma^\text{SM}}{\mathrm{d}\cos\theta})\right|_{\cos\theta=0} & = \frac{8-16\sin^2\theta_W+8\sin^4\theta_W}{9-16\sin^2\theta_W+8\sin^4\theta_W} \approx 0.82
\end{align}
in the same limit $\sqrt{s} \gg M_W$. The dashed horizontal line in Figs.~\ref{fig9} and \ref{fig10} represents this minimal value. It is worth noting that if initial $e^+$ and $e^-$ are unpolarized, then the coefficient of $\sin^4\theta_W$ in~\eqref{eq:seesaw_min} would be different (12 instead of 8), leading to a slight change in the minimum value.

From Eqs.~\eqref{eq:newmin} and~\eqref{eq:seesaw_min}, we see that if the cross section is measured with an accuracy at the level of 5~\% for energies up to $\sqrt{s}\sim M_Z/\abs{V_{eN}}$, it would be possible to see evidence for the existence of a heavy neutrino. If $M_N\gtrsim \sqrt{s}$, then we would reach the minimum and find the suppression of the cross section, followed by growth at higher energies. The only case where there is no sizable correction is if $M_N\lesssim \sqrt{s}\sim M_Z/\abs{V_{eN}}$, see condition~\eqref{first_cond}. 

\section{Constraints on $\abs{V_{\ell N}}^2$ at future lepton colliders}
\label{bounds}

In this section, we employ our \texttt{HeavyNU} model to estimate the prospective constraints on the mixing parameters\footnote{Since the muon colliders place constraints on $\abs{V_{\mu N}}^2$, we will plot $\abs{V_{\ell N}}^2$.} that can be achieved at future lepton colliders, and we compare them with the current limits obtained from LEPII data. We consider the following colliders: CEPC ($HZ$), FCC-ee ($t\bar{t}$), ILC, CLIC, BM1, and BM2. Our estimate is based on the method utilized in~\cite{Gabrielli:2026cjk}, which relies on evaluating a $\chi^2$-test defined by the deviation of the cross section in the presence of heavy neutrinos from the Standard Model prediction. In particular, for LEPII, this estimate for sufficiently large masses $M_N$ is based on six measurements of the total cross section at different center-of-mass energies [see~\cite{Gabrielli:2026cjk} for details]:
\begin{equation}
    \abs{V_{eN}}^2 \lesssim 0.0135.
\end{equation}
For future lepton colliders, we perform the analysis at a single fixed energy; consequently, the expression for the $\chi^2$ at 95\% C. L. takes the form~\cite{Gabrielli:2026cjk}:
\begin{equation}\label{chi_sq}
    \chi^2 = \frac{(N_\text{obs}-N_\text{exp})^2}{N_\text{exp}} = \mathcal{L}\frac{(\sigma_\mathcal{N}-\sigma_\text{SM})^2}{\sigma_\text{SM}} \leq 3.841.
\end{equation}
In this work, we will take into account only statistical errors. Systematic uncertainties (in particular, the uncertainty in the measurement of the momentum direction) are not expected to significantly affect the value of the integrated cross section since our integration region is much larger than the corresponding uncertainty. We extend the analysis provided in~\cite{Gabrielli:2026cjk} by explicitly accounting for the finite mass of the heavy neutrino $M_2$. 

In our analysis, we take into account the center-of-mass energies and integrated luminosities of all considered colliders. However, these colliders also differ in beam polarizations and angular acceptances. Since the latter two characteristics may change by the time the accelerators are constructed, we adopt a conservative approach in which all beams are assumed to be unpolarized, and a fixed angular acceptance $\theta\sim 10^\circ-170^\circ$ is imposed. The actual values of beam polarization and angular acceptance may either weaken or strengthen the results presented below.

\begin{figure}[p]
    \centering
    \begin{minipage}{1.0\textwidth}
        \centering
        \includegraphics[width=6in]{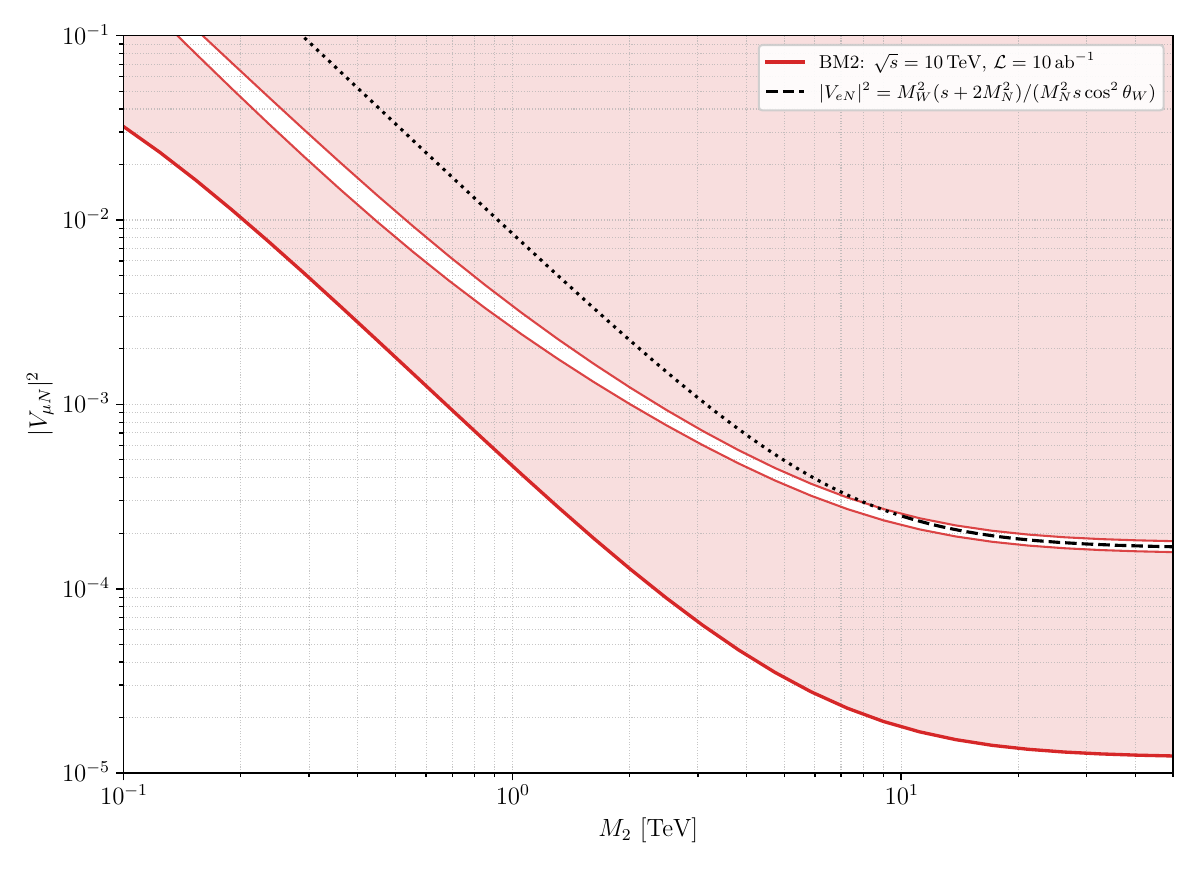}
        \caption{BM2 exclusion regions  in $(M_2,|V_{\mu N}|^2)$ parameter space for integration in $-\cos 10^\circ \leq \cos\theta \leq \cos 10^\circ$. Here, the black line represents formula~(\ref{eq:V_zero}) (dotted for $M_2 \lesssim \sqrt{s}$, dashed for \mbox{$M_2 \gtrsim \sqrt{s}$)}.}
        \label{BM2}
    \end{minipage}
    \vspace{0.02\textwidth}
    \begin{minipage}{1.0\textwidth}
        \centering
        \includegraphics[width=6in]{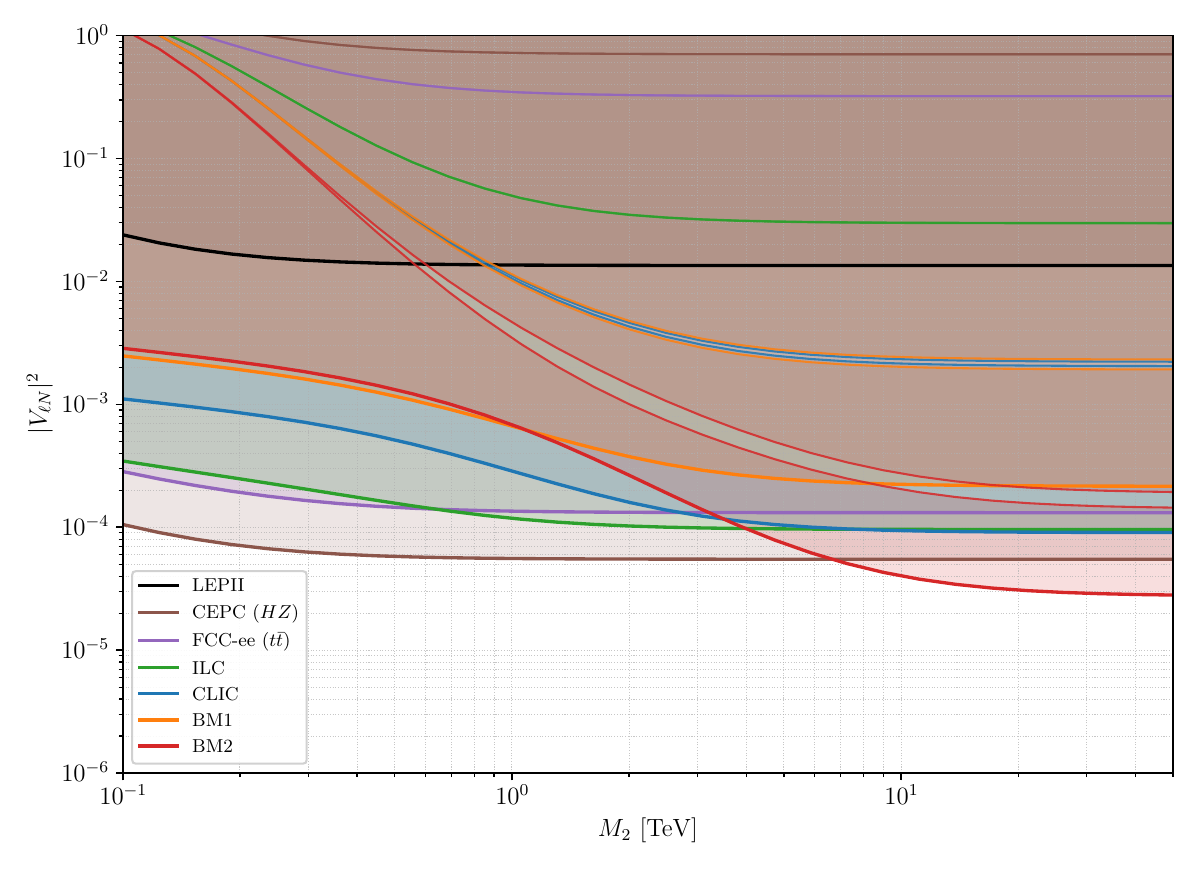}
        \caption{Exclusion regions in $(M_2,|V_{\ell N}|^2)$ parameter space for integration in $-1\leq \cos\theta \leq 1$.}
        \label{bound10}
    \end{minipage}
\end{figure}

The results of our analysis are presented as exclusion regions in the following figures. It is important to note that, in addition to the lower allowed region of the mixing parameter, a narrow upper allowed region also appears (see Fig.~\ref{BM2}). As discussed earlier (see Figs.~\ref{fig9}~and~\ref{fig10}), with increasing $\abs{V_{\ell N}}^2$, the differential cross section $\frac{\dd\sigma^\mathcal{N}}{\dd\cos\theta}$ first becomes smaller than the SM prediction; however, later it becomes larger. Consequently, for a certain nonzero value of $\abs{V_{\ell N}}^2$, the cross section $\sigma^\mathcal{N}$ coincides with the SM one, leading to a vanishing $\chi^2$. Therefore, in the vicinity of this value, the $\chi^2$-condition is satisfied, and the corresponding region of the mixing parameter is not excluded. These regions correspond precisely to the narrow bands of weaker constraints obtained for the colliders. 

\begin{figure}[t]
    \centering
    \includegraphics[width=6in]{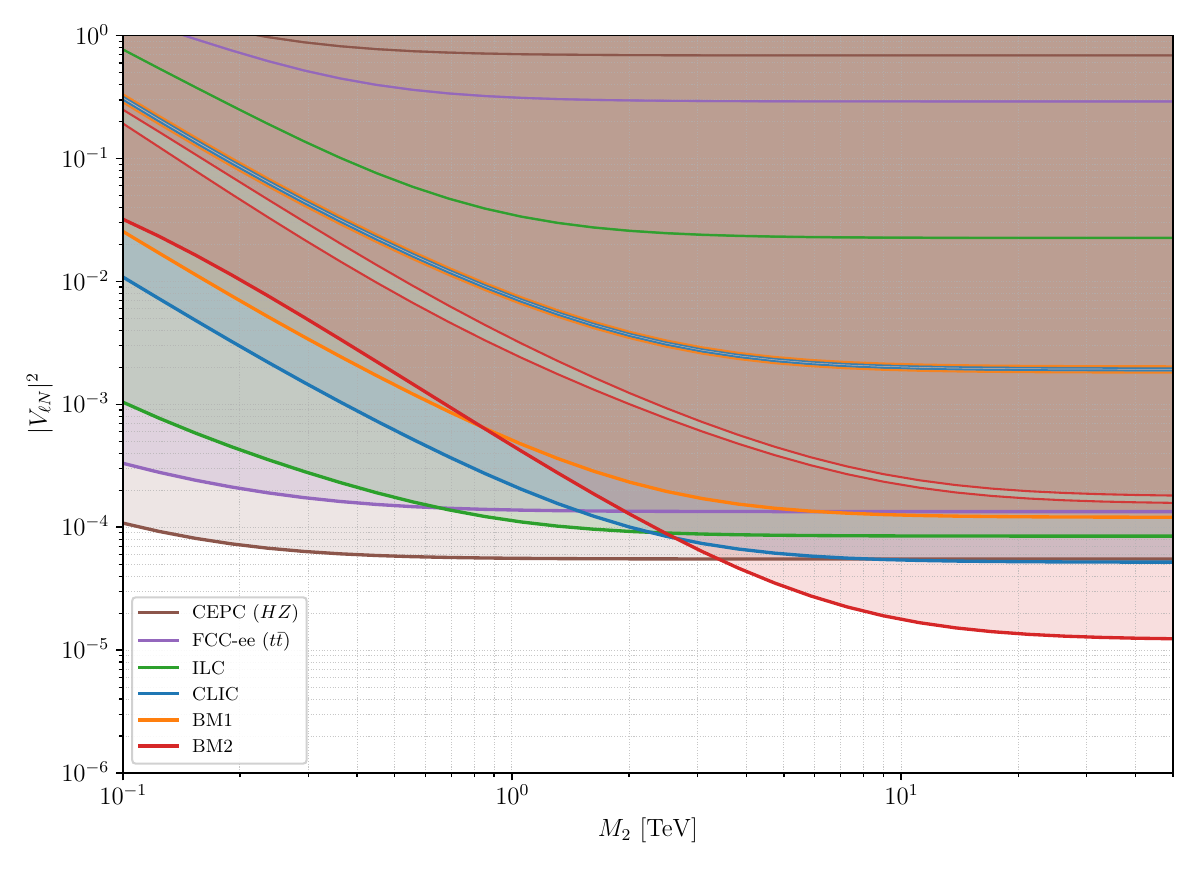}
    \caption{Exclusion regions in $(M_2,|V_{\ell N}|^2)$ parameter space for integration in $-\cos 10^\circ\leq \cos\theta \leq \cos 10^\circ$.}
    \label{bound098}
\end{figure}

\begin{figure}[t]
    \centering
    \includegraphics[width=6in]{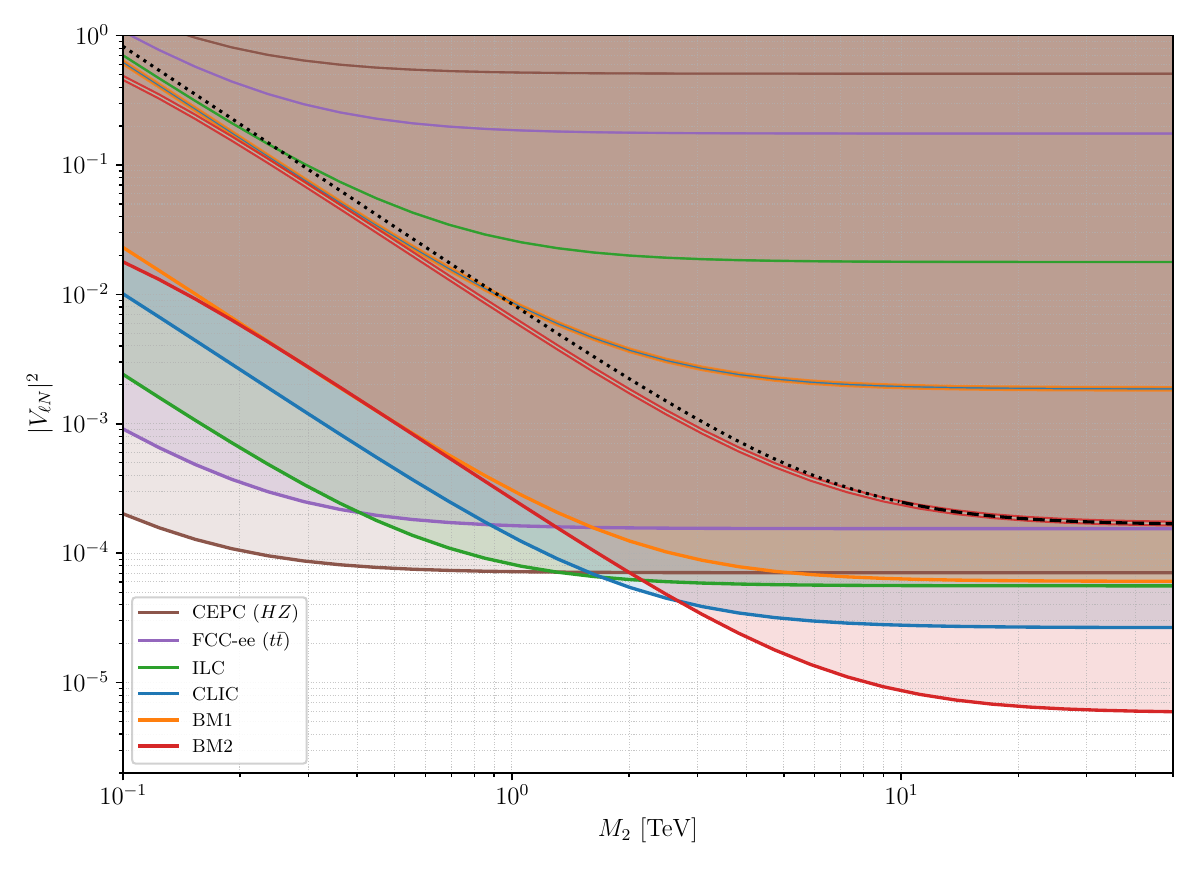}
    \caption{Exclusion regions in $(M_2,|V_{\ell N}|^2)$ parameter space for integration in $-0.6\leq \cos\theta \leq 0.6$. Here, the black line represents the formula~(\ref{eq:V_zero}) with $\sqrt{s} = 10$ TeV (dotted for $M_2 \lesssim \sqrt{s}$, dashed for \mbox{$M_2 \gtrsim \sqrt{s}$)}.}
    \label{bound06}
\end{figure}

This nonzero value of the mixing parameter can be estimated using the result for the minimum of the differential cross section [see~(\ref{eq:newmin})]. Since the curves in Figs.~\ref{fig9}~and~\ref{fig10} are symmetric about their minimum, the value of the mixing parameter that vanishes $\chi^2$ is twice as large as the one from~(\ref{eq:newmin}):
\begin{equation}
   \label{eq:V_zero}
    |V_{\ell N}|^2_{\sigma^\mathcal{N}=\sigma^\text{SM}} = \frac{M_W^2\qty(s+2M_N^2)}{M_N^2s\cdot \cos^2\theta_W}.
\end{equation}
Figure~\ref{BM2} represents this result for the muon collider BM2. We can see that, for large values of $M_N \gtrsim \sqrt{s}$, formula~(\ref{eq:V_zero}) can be used to predict the location of the upper allowed region. However, for smaller masses $M_N \lesssim \sqrt{s}$, this formula becomes inaccurate. This occurs because, at smaller masses, the impact of the noncentral angular region becomes important, while formula~(\ref{eq:V_zero}) is only relevant for $\cos\theta = 0$. However, when the angular integration range is restricted, formula~(\ref{eq:V_zero}) becomes more accurate (see Fig.~\ref{bound06}). 

We start with the total cross sections ($-1 \leq \cos\theta \leq 1$). From Fig.~\ref{bound10}, it can be seen that a second allowed region exists for almost all considered accelerators. An exception is provided by LEPII, for which only a single allowed region exists. This is because, in the LEPII case, $\chi^2$ includes several measurements performed at different center-of-mass energies, and for each energy, the cross section vanishes at a different value of the mixing parameter. As a result, $\chi^2$ does not vanish for any value of $\abs{V_{\ell N}}^2$, and no second allowed region emerges. In principle, if future colliders operate with stepwise increases in energy, measurements of the relevant cross sections at multiple energies would likewise allow one to exclude the upper allowed region. In the further analysis, we will focus on the lower bound on $\abs{V_{\ell N}}^2$.

From Fig.~\ref{bound10}, it is evident that, for heavy neutrino masses $M_2 \gtrsim \sqrt{s}$, the lower bound becomes independent of $M_2$, reproducing\footnote{Let us note that there is no upper allowed region in~\cite{Gabrielli:2026cjk}.} the results of~\cite{Gabrielli:2026cjk}. However, in the intermediate mass region $M_2 \gtrsim M_W$, the lower constraints become weaker, although they remain significantly stronger than those obtained at LEPII. 

We note that, in the regime $M_2 \gtrsim \sqrt{s}$, the constraints can be further improved by restricting the angular integration range. As pointed out in~\cite{Drutskoy:2024hqx}, the SM cross section is dominated by a narrow forward peak near $\cos\theta =1$, whereas the contribution from sufficiently heavy neutrinos is maximal in the central region~\mbox{$\cos\theta \sim 0$}. This suggests that a comparison of cross sections in the central angular region is more sensitive to New Physics effects. In the $\chi^2$-analysis, the forward peaks largely cancel in the numerator of~(\ref{chi_sq}), leaving the difference governed primarily by the heavy neutrinos' contribution to the process, while the SM peak contributes dominantly to the denominator, enhancing the total cross section. By excluding the forward region, the numerator remains controlled solely by the heavy neutrinos' contribution, whereas the denominator is significantly reduced, leading to stronger bounds on $\abs{V_{\ell N}}^2$.

\begin{figure}[p]
    \centering
    \begin{minipage}{1.0\textwidth}
        \centering
        \includegraphics[width=6in]{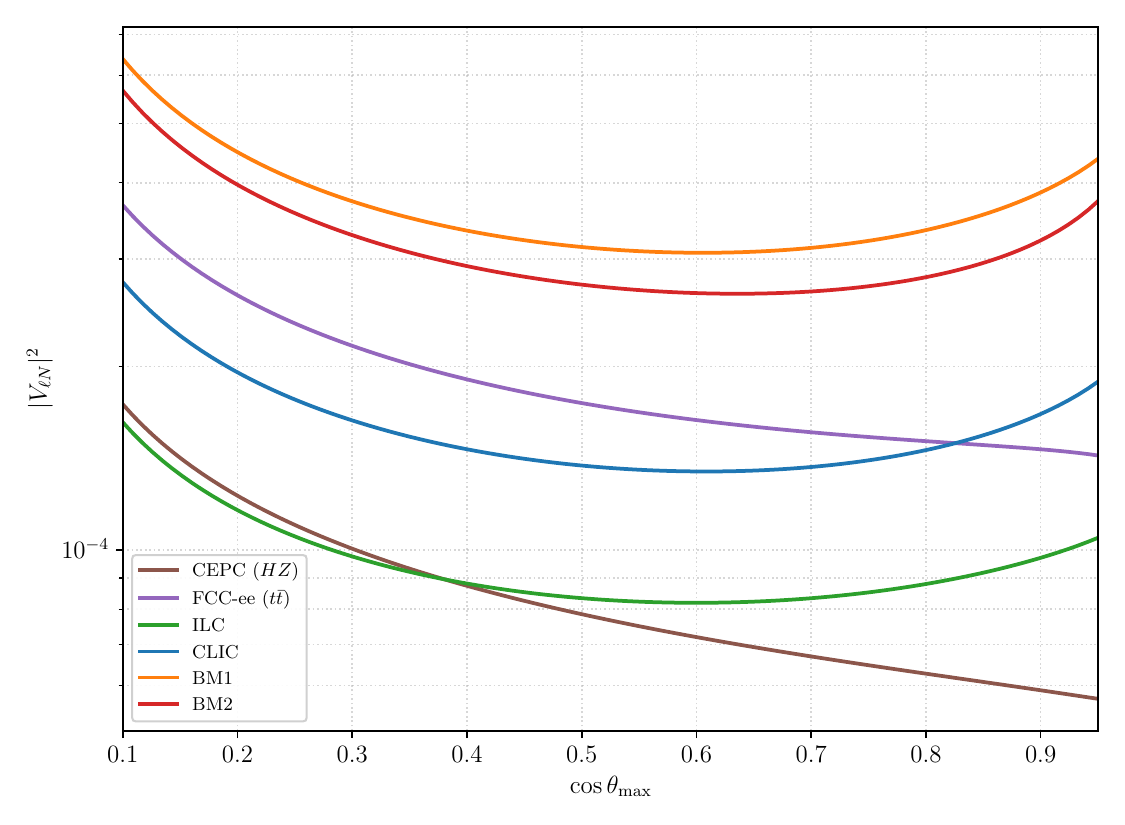}
        \caption{Dependence of the lower bound on the mixing parameter $|V_{\ell N}|^2$ on the angular region $-\cos\theta_\text{max}\leq \cos\theta \leq \cos\theta_\text{max}$ for $M_2 = 1$~TeV.}
        \label{cos_max1}
    \end{minipage}
    \vspace{0.02\textwidth}
    \begin{minipage}{1.0\textwidth}
        \centering
        \includegraphics[width=6in]{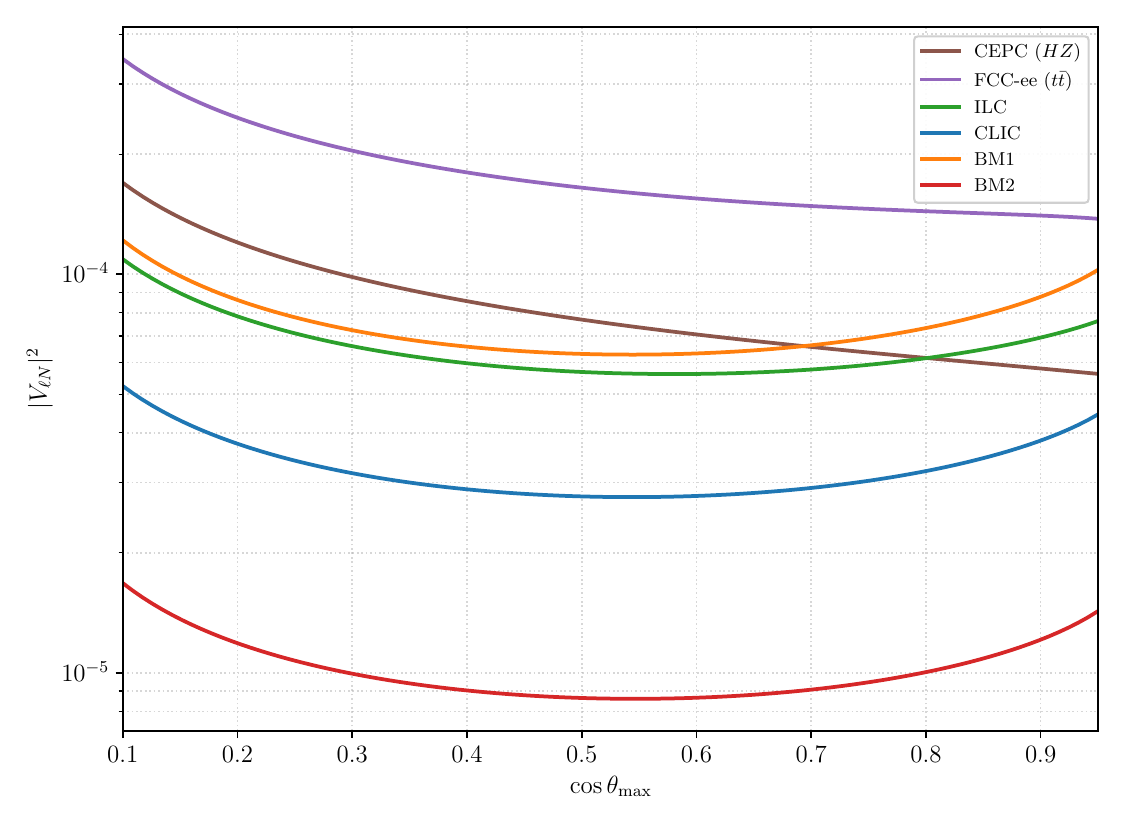}
        \caption{Dependence of the lower bound on the mixing parameter $|V_{\ell N}|^2$ on the angular region $-\cos\theta_\text{max}\leq \cos\theta \leq \cos\theta_\text{max}$ for $M_2 = 10$~TeV.}
        \label{cos_max10}
    \end{minipage}
\end{figure}

The corresponding results are shown in Figs.~\ref{bound098}~and~\ref{bound06}. The former figure represents the cuts that occur due to the angular acceptance under consideration. As we can see, the bound in the region $M_2 \gtrsim \sqrt{s}$ has already become stronger, which means that the forward peak no longer contributes. A cut at $\cos\theta =0.6$ yields the strongest constraint (see Figs.~\ref{cos_max1}~and~\ref{cos_max10}). At the same time, the asymptotic behavior of the upper constraints in the region $M_2 \lesssim \sqrt{s}$ approaches the same asymptotic regime $\abs{V_{\ell N}}^2 = \frac{M_W^2}{M_2^2\cos^2\theta_W}$ for all considered colliders as formula~(\ref{eq:V_zero}) works much better. Let us also note that this optimal choice of angle roughly corresponds to the barrel part of the detectors. The angular region is chosen symmetrically to ensure applicability in scenarios where the charge of the reconstructed $W$-bosons cannot be determined~\cite{Drutskoy:2024hqx}. Importantly, this symmetrization does not lead to a substantial loss of statistics: integrating over $-0.6 \leq \cos\theta \leq 0.6$ reduces the number of events from the SM by only about 5\% compared to integration over $-1 \leq \cos\theta \leq 0.6$. As shown in Figs.~\ref{bound10}~and~\ref{bound06}, this procedure improves the constraints on the mixing parameter by more than an order of magnitude in the region $M_2 \gtrsim \sqrt{s}$ for almost all considered colliders. 

On the other hand, for moderately heavy neutrino masses $M_N\gtrsim M_W$, imposing such angular cuts significantly weakens the constraints (see, e. g.,~Figs.~\ref{bound10}~and~\ref{bound06}). This behavior can be understood in the following way. When the heavy neutrino mass $M_2$ becomes significantly smaller than the energy of the colliding particles, neutrino $N_2$ starts to behave like a light neutrino. This means that if we remove the forward region, we will eliminate not only the enhancement that comes from light neutrinos but also the modification that comes from heavy neutrinos.  

\begin{figure}[p]
    \centering
    \begin{minipage}{1.0\textwidth}
        \centering
        \includegraphics[width=6in]{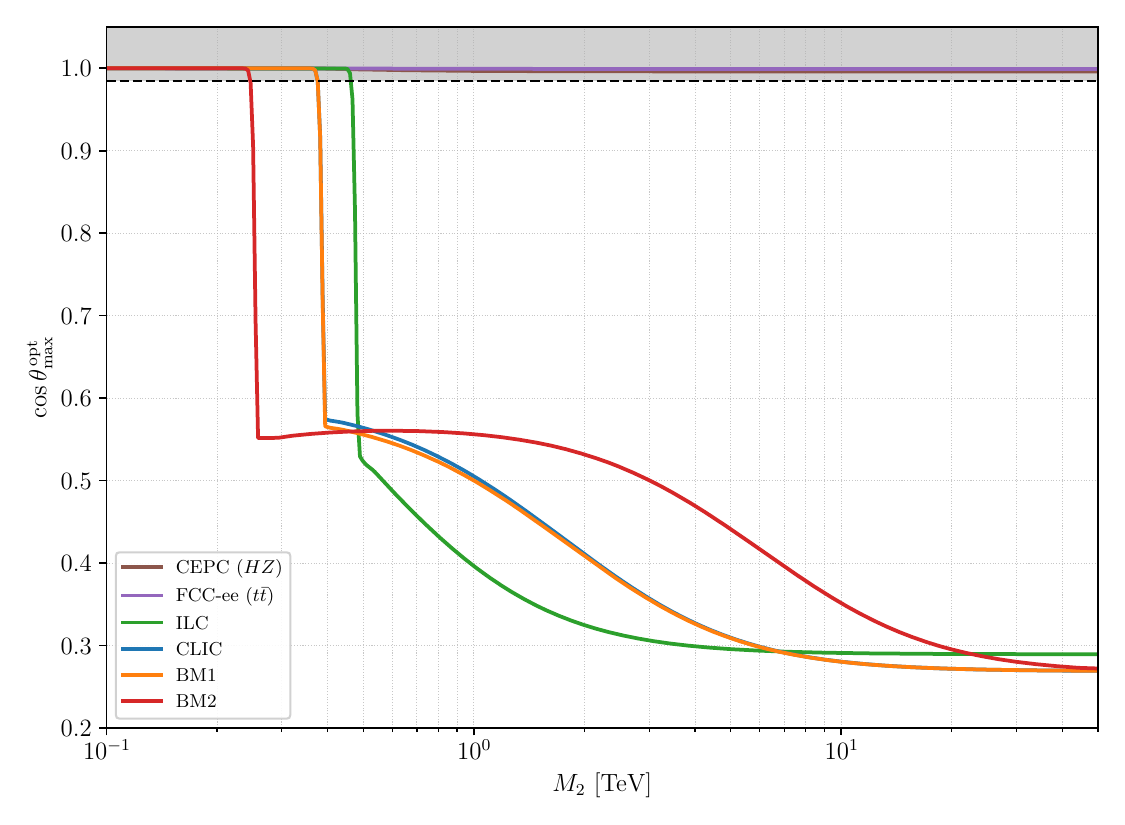}
        \caption{Dependence of the optimal value of $\cos\theta_\text{max}$ on heavy neutrino mass $M_2$ for the nonsymmetric angular region $-1\leq \cos\theta \leq \cos\theta_\text{max}$. Here, the lines for CEPC ($HZ$) and FCC-ee ($t\bar{t}$) almost coincide, and the dashed line shows the angular acceptance region $\cos\theta = \cos 10^\circ$.}
        \label{opt_non-sim}
    \end{minipage}
    \vspace{0.02\textwidth}
    \begin{minipage}{1.0\textwidth}
        \centering
        \includegraphics[width=6in]{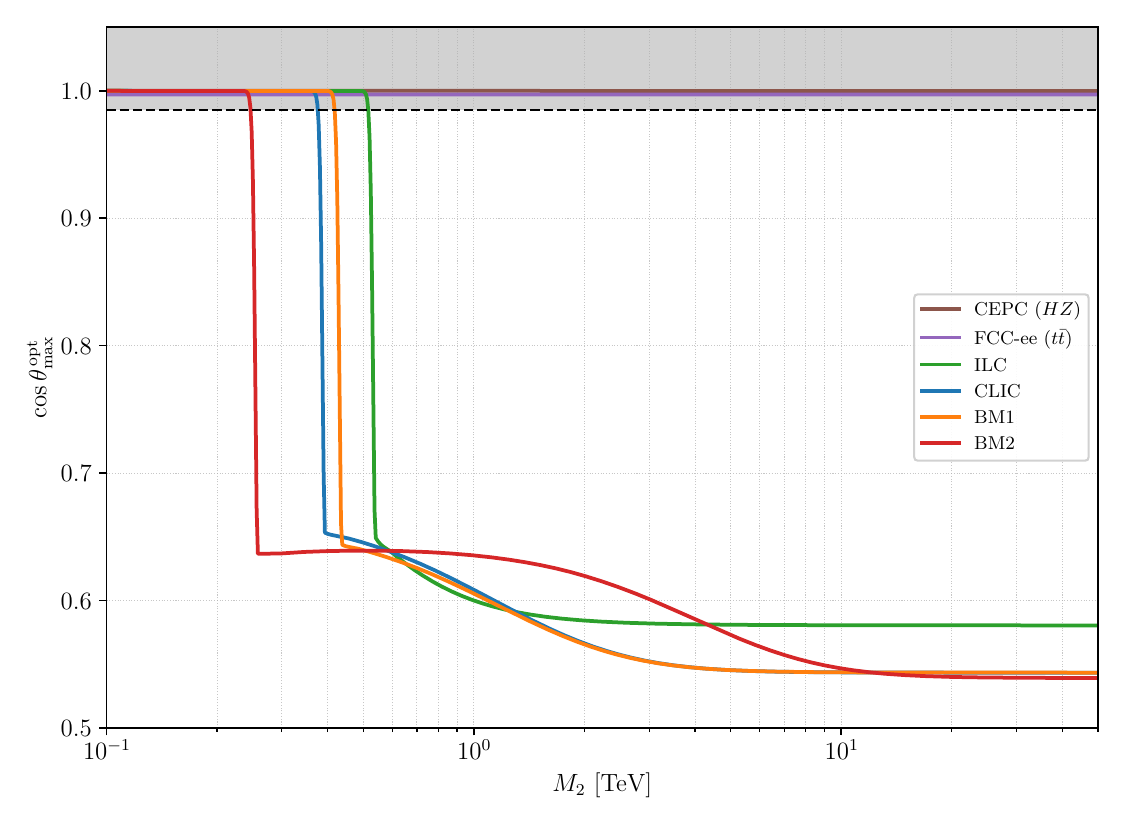}
        \caption{Dependence of the optimal value of $\cos\theta_\text{max}$ on heavy neutrino mass $M_2$ for the symmetric angular region $-\cos\theta_\text{max}\leq \cos\theta \leq \cos\theta_\text{max}$. Here, the lines for CEPC ($HZ$) and FCC-ee ($t\bar{t}$) almost coincide, and the dashed line shows the angular acceptance region $\cos\theta = \cos 10^\circ$.}
        \label{opt_sim}
    \end{minipage}
\end{figure}

Figures~\ref{cos_max1}~and~\ref{cos_max10} illustrate how the lower constraint on $\abs{V_{\ell N}}^2$ depends on the angular integration range $-\cos\theta_\text{max} \leq \cos\theta \leq \cos\theta_\text{max}$. For all considered colliders, there exists an optimal value of $\cos\theta_\text{max}$ at which the bound on $\abs{V_{\ell N}}^2$ is minimized. However, the optimal value for CEPC~($HZ$) and FCC-ee~($t\bar{t}$) is $\cos\theta_\text{max}=1$; at such energies, the peak around $\cos\theta = 1$ is significantly broadened, and the contribution of the heavy neutrinos is noticeable even near $\cos\theta = 1$, which leads to an improved constraint when the peak region is included in the integration domain.

\begin{figure}[t]
    \centering
    \includegraphics[width=6in]{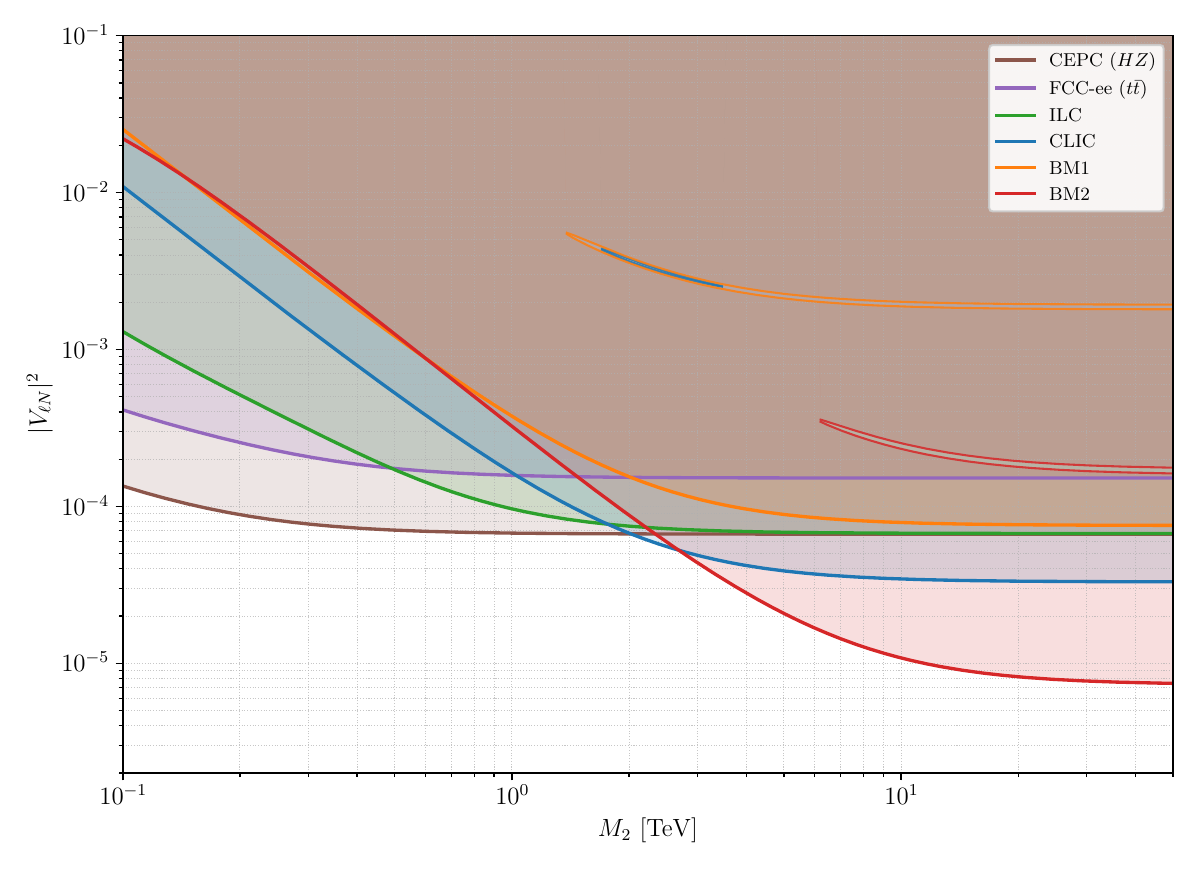}
    \caption{Exclusion regions in $(M_2,|V_{\ell N}|^2)$ parameter space in the case of two bins.}
    \label{2bins}
\end{figure}

Finally, we investigate the optimal choice of the angular integration region in more detail. Figures~\ref{opt_non-sim}~and~\ref{opt_sim} show the regions that yield the strongest bounds on the mixing parameter $\abs{V_{\ell N}}^2$ for different heavy neutrino masses. We find that, for sufficiently small $M_2$, the best sensitivity is achieved by integrating over the full angular range, whereas for larger masses, it is advantageous to restrict the analysis to the central region.

It is also possible to keep most of the advantages of both approaches. The angular integration region can be partitioned into bins, with the integrated cross section evaluated in each bin, followed by a $\chi^2$-test with the number of data points equal to the number of bins. In particular, we can consider two bins, $[0.0, 0.6]$ and $[0.6, \cos 10^\circ]$, which correspond to the central and forward regions, respectively. The results are shown in Fig.~\ref{2bins}. Comparing to Fig.~\ref{bound098}, we can state that the lower constraint on the mixing parameter $\abs{V_{\ell N}}^2$ has become more stringent for $M_2 \gtrsim 1$ TeV, while the upper constraint has disappeared for CEPC ($HZ$), FCC-ee ($t\bar{t}$), and ILC, and has remained only in specific mass regions for CLIC, BM1, and BM2. At the same time, if we compare Figs.~\ref{bound06}~and~\ref{2bins}, the lower constraints in the region $M_2 \gtrsim M_W$ have become more rigorous for the approach with two bins. This means that this approach helps both to strengthen the bounds in the region $M_2 \gtrsim M_W$ and to keep stringent constraints in the region $M_2 \gtrsim 1$ TeV. Let us also note that considering more than two bins does not lead to additional enhancement of the constraints.

Summing up the results, we can point out general features. Despite having the lowest center-of-mass energy among the colliders considered, CEPC or FCC-ee in $HZ$ mode provides one of the strongest constraints due to its high integrated luminosity. This indicates that precise cross section measurements at low energies are competitive to those at high energy colliders and can even surpass them in certain regions.

\section{Conclusions}
\label{concl}

In this work, we have studied the process $e^+e^- \to W^+W^-$ within a seesaw type-I framework. An analytic calculation for a single heavy neutrino was performed in a simplified toy model, the results of which reproduce the key features of the corresponding extension of the Standard Model. We have also obtained results for the case of three heavy neutrinos, for which we provide \texttt{FeynRules} model files.

Furthermore, we have highlighted the main differences between the exact unitary mixing of light and heavy neutrinos and its linearized approximation, the latter becoming inapplicable once the center-of-mass energy $\sqrt{s}$ reaches the point where the SM cross section crosses the asymptotic regime. At the same time, we have shown that, for $M_N \gtrsim \frac{M_W\sqrt{3}}{\abs{V_{\ell N}}}$, the exact unitary mixing leads to a noticeable enhancement in comparison to the Standard Model cross section starting from energies $\sqrt{s}\sim\frac{M_W\sqrt{3}}{\abs{V_{\ell N}}} $, while the linearized mixing remains almost indistinguishable from it, enhancing the cross section only at higher energies. In addition, when $\sqrt{s} \gtrsim M_N$, the exact unitary mixing retains the correct behavior, $\sigma\propto \frac{1}{s}$, and remains significantly modified in comparison with the SM cross section, while the linearized mixing grows indefinitely with center-of-mass energy and breaks $S$-matrix unitarity.

It is also important that the cross section can be suppressed by up to 18\% in the exact unitary mixing approach, while for the linearized mixing, the cross section is always greater than in the SM. This implies that precise measurements of the differential cross section in the central angular region could also reveal the presence of heavy neutrinos.

Taken together, these observations indicate that exact unitary mixing is not only the theoretically consistent way of incorporating heavy neutrinos, which preserves the unitarity of the $S$-matrix, but also the more accessible option from an experimental perspective. Finally, we have presented numerical results for a range of collider energies, which can be used to derive bounds on the heavy-light neutrino mixing parameters for arbitrarily large masses.

Furthermore, we performed an estimate of constraints that could possibly be obtained at future lepton colliders. These bounds can be compared to the ones obtained earlier~\cite{Atre:2009rg,CMS:2022hvh,Antusch:2015mia}. It is clear that direct searches can compete with the obtained results only for $M_N$ close to the electroweak scale, while constraints from EW PO are comparable to the obtained bounds [see Fig.~7 from~\cite{Antusch:2015mia}; the comparison, however, is complicated by different C. L. value]. 

Let us note that while this work was being completed, the paper \cite{Gabrielli:2026cjk} was published, reporting closely related results. In~\cite{Gabrielli:2026cjk}, the limit $M_N\to\infty$ was considered, so there is no dependence on $M_N$ anywhere. In our paper, we have found the conditions for $M_N$ and mixing parameters in order to obtain sizable corrections to the SM predictions. We also emphasize that the minimal value of the ratio to the SM differential cross section~\eqref{eq:seesaw_min} does not depend on both $\sqrt{s}$ and $M_N$ in the relevant parameter region. At the same time, we point out that if only one measurement of the cross section is performed, a second allowed region appears. The comparison with the linearized mixing is one of the main points in our paper, and it could not be done in the limit $M_N\to\infty$ since all effects would be at $\sqrt{s}\to\infty$.

\section*{Acknowledgments}

We sincerely thank M. I. Vysotsky, A. G. Drutskoy, and E. S. Vasenin for the valuable discussions and ideas.

\bibliographystyle{unsrturl}
\bibliography{ref}

\end{document}